\documentclass[aps,prd,reprint,showpacs,longbibliography,
titlepage,nofootinbib]{revtex4-1}

\usepackage{amsthm}
\usepackage{amssymb}   
\usepackage{dsfont}
\usepackage{mathtools} 
\usepackage{hyperref}
\hypersetup{colorlinks=true,linkcolor=blue,urlcolor=blue,citecolor=blue}
\usepackage{accents}
\usepackage{tensor}
\usepackage{xcolor}
\usepackage[cal=boondox]{mathalfa}
\usepackage{lipsum}
\usepackage{relsize}
\usepackage{color}
\usepackage{comment}
\usepackage{nameref}
\usepackage{anyfontsize}

\usepackage[T1]{fontenc}
\newcommand{\ovg}[1]{\stackrel{(\gamma)}{#1}}
\newcommand{\uac}[1]{\underaccent{\tilde}{#1}}

\begin{document}
	
	\title{Hamiltonian analysis of fermions coupled to the Holst action}
	
	\author{Jorge Romero\href{https://orcid.org/0000-0001-8258-6647} {\includegraphics[scale=0.05]{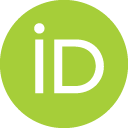}}}
	\email{ljromero@fis.cinvestav.mx}
	\author{Merced Montesinos\href{https://orcid.org/0000-0002-4936-9170} {\includegraphics[scale=0.05]{ORCIDiD_icon128x128.png}}}
	\email[Corresponding author \\]{merced@fis.cinvestav.mx}
	
	\affiliation{Departamento de F\'{i}sica, Centro de Investigaci\'on y de Estudios Avanzados del Instituto Politécnico Nacional, Avenida Instituto Polit\'{e}cnico Nacional 2508,\\
		San Pedro Zacatenco, 07360 Gustavo Adolfo Madero, Ciudad de M\'exico, Mexico}
	
	\author{Mariano Celada\href{https://orcid.org/0000-0002-3519-4736} {\includegraphics[scale=0.05]{ORCIDiD_icon128x128.png}}}
	\email[]{mcelada@matmor.unam.mx}
	\affiliation{Centro de Ciencias Matem\'{a}ticas, Universidad Nacional Aut\'{o}noma de M\'{e}xico,\\
		UNAM-Campus Morelia, Apartado Postal 61-3, Morelia, Michoac\'{a}n 58090, Mexico}
	
	\date{\today}
	
	\begin{abstract}
		We report three manifestly Lorentz-invariant Hamiltonian formulations of minimally and nonminimally coupled fermion fields to the Holst action. These formulations are achieved by making a suitable parametrization of both the tetrad and the Lorentz connection, which allows us to integrate out some auxiliary fields {\it without} spoiling the local Lorentz symmetry. They have the peculiarity that their noncanonical symplectic structures as well as the phase-space variables for the gravitational sector are real. Moreover, two of these Hamiltonian formulations involve half-densitized fermion fields. We also impose the time gauge on these formulations, which leads to real connections for the gravitational configuration variables. Finally, we perform a symplectomorphism in one of the manifestly Lorentz-invariant Hamiltonian formulations and analyze the resulting formulation, which becomes the Hamiltonian formulation of fermion fields minimally coupled to the Palatini action for particular values of the coupling parameters.  
	\end{abstract}
	
	\maketitle
	
	\section{Introduction}\label{intro}
	In the first-order formalism, the gravitational variables are an orthonormal frame of 1-forms $e^I$ and a Lorentz connection 1-form $\omega^I{}_J$. Because of its covariant nature under local Lorentz transformations, fermions couple naturally to the gravitational field in this framework~\cite{Weyl_1950,Kibble1,Kibble2}. In particular, in the self-dual approach to the coupling of fermion fields to general relativity, the full local Lorentz symmetry is preserved, but at the Hamiltonian level the theory involves the complex-valued Ashtekar connection as the configuration variable for general relativity that is difficult to deal with at the quantum level~\cite{Jacobson_1988,ART}. On the other hand, in the real case, taking the Holst action~\cite{Holst9605} to describe the gravitational field and adding to it the fermionic Lagrangian, the resulting first-order theory is related to the Einstein-Dirac theory supplemented with both interaction and boundary terms that depend on the nature of the coupling between fermions and gravity once the connection is integrated out in the action principle~\cite{Perez0602,Mercuri0604,Freidel0511,Alexandrov0806}.
	
	It is clear that to better understand the nature of the gravity-fermion interaction, the local Lorentz symmetry must be preserved in the Hamiltonian analysis too. However, some of the canonical approaches involving a real connection variable for the gravitational field break local Lorentz symmetry down to $SU(2)$ from the very beginning by imposing the time gauge~\cite{Thiemann_1998_1487,Bojowald0809}. Therefore, it is indispensable to perform a manifestly Lorentz-invariant Hamiltonian analysis of fermion fields coupled to general relativity that at the same time keeps the configuration variable of the gravitational field real. This would allow us to appreciate in depth the nature of the coupling between fermion and gravitational fields. Fortunately, a Hamiltonian analysis preserving Lorentz invariance for the Holst action was carried out recently~\cite{Montesinos2004a}, and the approach of such a paper will be taken here as the theoretical tool to study the coupling of fermion fields to general relativity. Using this approach, we report in Secs.~\ref{Sec_Ham_form}--\ref{Sec_hf2} three manifestly Lorentz-invariant Hamiltonian formulations of minimally and nonminimally coupled fermions to the Holst action. In any of the three cases, the symplectic structure is real and the phase-space variables that correspond to the gravitational sector are also real. In particular, the Hamiltonian formulations contained in Secs.~\ref{Sec_hf} and~\ref{Sec_hf2} involve half-densitized fermion fields. Furthermore, to compare our results with previous works on the subject, we impose the time gauge on these Hamiltonian formulations and break the local Lorentz symmetry down to $SO(3)$ [and thus to its double cover $SU(2)$] and we give the corresponding Hamiltonian formulations in Sec.~\ref{Sec_tg}, which involve real connections. Finally, we perform a symplectomorphism from the manifestly Lorentz-invariant Hamiltonian formulation of Sec.~\ref{Sec_hf2} and report the resulting Hamiltonian formulation in Sec.~\ref{Sec_CT}. We also analyze how this latter formulation simplifies when two particular values of the coupling parameters are chosen. Remarkably, one of these manifestly Lorentz-invariant Hamiltonian formulations corresponds to fermion fields minimally coupled to the Palatini action (Einstein-Cartan-Dirac). The other corresponds to fermion fields nonminimally coupled to the Palatini action, but it is also invariant under parity transformations.         
	
	\subsection{Conventions}
	In the first-order formalism, the fundamental variables for general relativity are an orthonormal frame of $1$-forms $e^I$ and a connection 1-form $\omega^I{}_J$ compatible with the metric $(\eta_{IJ}) =\mbox{diag} (-1, 1, 1, 1)$, $d \eta_{IJ} - \omega^K{}_I \eta_{KJ} - \omega^K{}_J  \eta_{IK}=0$, and therefore $\omega_{IJ}=-\omega_{JI}$ because Lorentz indices $I,J,K, \ldots$ (taking the values $0, 1, 2, 3$) are raised and lowered with $\eta_{IJ}$. The totally antisymmetric Lorentz-invariant tensor $\epsilon_{IJKL}$ is such that $\epsilon_{0123}=1$.  The antisymmetric part of tensors involving Lorentz indices is defined by $t_{[IJ]}= \left (t_{IJ} - t_{JI} \right )/2$. Furthermore, for any antisymmetric object $U_{IJ}=-U_{JI}$ we define its internal dual as $\ast U_{IJ} := (1/2) \epsilon_{IJKL}U^{KL}$ and also the object $\stackrel{(\gamma)}{U} _{IJ}:=P_{IJKL}U^{KL}$ with
	\begin{equation}
		\label{P_def}
		P^{IJ}{}_{KL} :=  \delta^{I}_{[K} \delta^{J}_{L]} + \frac{1}{2 \gamma} \epsilon^{IJ}{}_{KL},
	\end{equation}
	where $\gamma$ is the nonvanishing real Immirzi parameter. The weight of a tensor is either indicated with tildes over or below it, or mentioned explicitly. In particular, the spacetime tensor density $\underaccent{\tilde}{\eta}_{\mu\nu\lambda\sigma}$ ($\tilde{\eta}^{\mu\nu\lambda\sigma}$) is totally antisymmetric and such that $\underaccent{\tilde}{\eta}_{t123}=1$ ($\tilde{\eta}^{t123} =1)$. In addition, we define the three-dimensional Levi-Civita symbols as $\underaccent{\tilde}{\eta}_{abc}:=\underaccent{\tilde}{\eta}_{tabc}$ ($\tilde{\eta}^{abc}:=\tilde{\eta}^{tabc}$) and $\epsilon_{ijk}:=\epsilon_{0ijk}$. We assume that the spacetime $M$ has the topology $\mathbb{R}\times\Sigma$ and that $\Sigma$ has no boundary. We foliate $M$ by constant time hypersurfaces $\Sigma_t$ diffeomorphic to $\Sigma$. The coordinate $x^a$ labels the points of $\Sigma$ and $t$ labels the points of $\mathbb{R}$.

	\section{First-order action principle}\label{Sec_eff_action}
	We begin our analysis by considering for gravity the Holst action~\cite{Holst9605} with a cosmological constant $\Lambda$, given by
	\begin{equation}
		\label{S_H}
		S_{H}[e, \omega] = \kappa \int_{M} \left[ P^{IJ}{}_{KL} \ast \left( e^{K} \wedge e^{L} \right) \wedge F_{IJ} - 2 \Lambda \rho \right],
	\end{equation}
	where $F^{I}{}_{J} := d \omega^{I}{}_{J} + \omega^{I}{}_{K} \wedge \omega^{K}{}_{J}$ is the curvature of the connection $\omega^{I}{}_{J}$, $\rho:= (1/4!) \epsilon_{IJKL} e^{I} \wedge e^{J} \wedge e^{K} \wedge e^{L} $ is the volume form, and $\kappa = (16 \pi G)^{-1}$ with $G$ being Newton's gravitational constant.
	
	On the other hand, we consider the coupling of a Grassmann-valued fermion field $\psi$ to the gravitational field through the real fermionic action
	\begin{eqnarray}
		\label{S_F}
		S_{F}[e, \omega, \psi, \bar{\psi}] &:=& \int_{M} \bigg\lbrace  \frac{1}{2} \Big[ \bar{\psi} \gamma^{I} E D \psi - \overline{D \psi} \gamma^{I} E^{\dagger} \psi \Big] \wedge \star e_{I} \notag \\
		& & - \chi m \bar{\psi} \psi \rho \bigg\rbrace,
	\end{eqnarray}
	where $m$ is the mass of $\psi$, $\bar{\psi}:= \mathrm{i} \psi^{\dagger} \gamma^{0}$ ($\mathrm{i}$ is the imaginary unit), $\star$ denotes the Hodge dual [and so $\star e_{I} = (1/3!)\epsilon_{IJKL} e^{J} \wedge e^{K} \wedge e^{L}$], $\gamma^{I}$ denotes Dirac's matrices [and so they satisfy $ \gamma^{I} \gamma^{J} +  \gamma^{J} \gamma^{I} = 2 \eta^{IJ} \mathds{1}$ and $(\gamma^{I})^{\dagger} = \gamma^{0} \gamma^{I} \gamma^{0}$], and the covariant derivatives of $\psi$ and $\bar{\psi}$ are defined by 
	\begin{subequations}
		\begin{eqnarray}
			D \psi &:= & d \psi + \frac{1}{2} \omega_{IJ} \sigma^{IJ} \psi, \\ 
			\overline{D \psi} &:= &  d \bar{\psi} - \frac{1}{2} \omega_{IJ} \bar{\psi} \sigma^{IJ},
		\end{eqnarray}
	\end{subequations}
	with $\sigma^{IJ} := (1/4) [\gamma^{I}, \gamma^{J}]$ being the Lorentz generators in the spin representation. Note that in~\eqref{S_F} $E:= \left(\chi + \mathrm{i} \theta\right) \mathds{1} - \mathrm{i} \xi \gamma^{5}$ denotes the coupling matrix with $\gamma^{5}:= \mathrm{i} \gamma^{0} \gamma^{1} \gamma^{2} \gamma^{3}$ and $\chi$, $\theta$, and $\xi$ being real adimensional parameters. As a consequence, $E^{\dagger}= \left(\chi - \mathrm{i} \theta\right) \mathds{1} + \mathrm{i} \xi \gamma^{5}$ since $\gamma^{5}$ is a Hermitian matrix. Also, note that $\gamma^{5}$ anticommutes with $\gamma^I$ ($\gamma^{5} \gamma^{I} = - \gamma^{I} \gamma^{5}$). Before going on, let us comment on the parameters $\chi$, $\theta$ and $\xi$. At this stage, we note that it is possible to make a redefinition of the fermion field $\psi \mapsto \chi^{-1/2} \psi$ and $\bar{\psi} \mapsto \chi^{-1/2} \bar{\psi}$ in the action~\eqref{S_F} that transforms $E$ into $E:= \left(1 + \mathrm{i} {\theta} \chi^{-1} \right) \mathds{1} - \mathrm{i} {\xi} \chi^{-1} \gamma^{5}$, which means that theory is really described by two parameters $\theta/\chi$ and $\xi/\chi$. This field redefinition is fixed by setting $\chi=1$, which we do now. Thus, $E:= \left(1 + \mathrm{i} {\theta} \right) \mathds{1} - \mathrm{i} {\xi} \gamma^{5}$ from now on.
	
	Therefore, the nonminimal coupling of a fermion field to general relativity in the first-order formalism is given by the real action
	\begin{equation}
		\label{S}
		S[e, \omega, \psi, \bar{\psi}] := S_{H}[e, \omega]  + S_{F}[e, \omega, \psi, \bar{\psi}].
	\end{equation}
	This action generalizes some of the cases already reported in the literature. For example, we recover the minimal coupling of fermion fields studied in Ref.~\cite{Perez0602} when $\theta=0=\xi$, whereas the nonminimal coupling considered in Ref.~\cite{Mercuri0604} or~\cite{Freidel0511} is obtained when $\theta=0$ (and $\xi \neq 0)$ and $\xi=0$ (and $\theta \neq 0$), respectively. We do not consider the action of Ref.~\cite{Alexandrov0806} because we have included a mass term in our analysis, and this requires an action principle that gives Dirac's equation when gravity is turned off~\cite{Kazmierczak0903} (see also Sec.~\ref{EM} of this paper).
	
	\subsection{Equations of motion}
	\label{EM}
	Computing the variation of the action~\eqref{S} with respect to $e^I$, $\omega^I{}_J$, $\psi$, and $\bar{\psi}$, we get the corresponding equations of motion which are as follows:
	\begin{subequations}
		\begin{eqnarray}
			\label{eom_e}
			&& \quad 2 \kappa {P\ast}{}^{KL}{}_{IJ} e^{J} \wedge F_{KL} - \left(2 \kappa \Lambda  + m \bar{\psi} \psi \right)  \star e_{I}  \notag \\
			&& + \dfrac{1}{2} \left(\bar{\psi} \gamma^{J} E D \psi - \overline{D \psi} \gamma^{J} E^{\dagger} \psi\right) \wedge \ast \left(e_{I} \wedge e_{J} \right) =0,\\
			\label{eom_w}
			& & \kappa P^{IJ}{}_{KL} D \ast \left( e^{K} \wedge e^{L} \right) - \dfrac{1}{4}  \epsilon^{IJ}{}_{KL}  A^{[K}  \star e^{L]} \notag \\
			& & - \dfrac{1}{2} \left( \xi A^{[I} + \theta V^{[I}\right) \star e^{J]} = 0, \\
			\label{eom_p}
			& &  \overline{D \psi} \gamma^{I} \wedge \star e_{I} + \dfrac{1}{2} \bar{\psi} \gamma^{I} E D \star e_{I} +  m \bar{\psi} \rho =0, \\
			\label{eom_bp}
			& & \gamma^{I} D \psi \wedge \star e_{I} +  \dfrac{1}{2} \gamma^{I} E^{\dagger} \psi D \star e_{I} -  m \psi \rho =0.
		\end{eqnarray}
	\end{subequations}
	Note that $ {P\ast}{}^{IJ}{}_{KL} := P^{IJ}{}_{MN} \tfrac{1}{2} \epsilon_{KL}{}^{MN}= P_{MN}{}^{IJ} \tfrac{1}{2} \epsilon_{KL}{}^{MN} =: {\ast P}_{KL}{}^{IJ}$. Here $D$ stands for the covariant derivative with respect to the connection $\omega^{I}{}_{J}$ and we have defined the vector and axial real currents, respectively, by
	\begin{eqnarray}
		\label{Vc_def}
		V^{I} &:=& i \bar{\psi} \gamma^{I} \psi, \\
		\label{Ac_def}
		A^{I} &:=& i \bar{\psi} \gamma^{5} \gamma^{I} \psi.
	\end{eqnarray}
	
	We now express the equations of motion in an equivalent form by solving~\eqref{eom_w} for $\omega^I{}_J$ and then by substituting the result in the remaining equations of motion. To do this, note that~\eqref{eom_w} can be cast in the form
	\begin{eqnarray}\label{FCE}
		D e^I = T^I,
	\end{eqnarray}
	where $T^I$ is the torsion of $\omega^I{}_J$,
	\begin{eqnarray}
		&& T^I = \dfrac{\gamma^2}{8 \kappa (\gamma^2 + 1)}\left[  \theta V_J + \left( \xi + \frac{1}{\gamma} \right) A_J \right] e^I \wedge e^J \nonumber \\
		&& + \dfrac{\gamma^2}{4 \kappa (\gamma^2 + 1)} \left[ \dfrac{\theta}{\gamma} V_J + \left( \dfrac{\xi}{\gamma} - 1 \right) A_J \right] \ast\left ( e^I \wedge e^J \right ).
	\end{eqnarray}
	The expression for $\omega^I{}_J$ that solves~\eqref{FCE} is 
	\begin{equation}
		\label{w_split}
		\omega^I{}_J = \Omega^I{}_J + C^I{}_J,
	\end{equation}
	where $\Omega^I{}_J$ is the spin connection defined by $\mathfrak{D} e^{I} := d e^{I} + \Omega^I{}_J \wedge e^{J} = 0$ and $d \eta_{IJ} - \Omega^K{}_I \eta_{KJ} - \Omega^K{}_J \eta_{IK}=0$ (and so $\Omega_{IJ}= -\Omega_{JI}$), while $C^I{}_J$ ($C_{IJ}=-C_{JI}$) is the contorsion 1-form defined by $T^{I} =: C^I{}_J \wedge e^J$ and given by
	\begin{eqnarray}
		C^I{}_J & = &  - \dfrac{1}{4 \kappa} \left( P^{-1} \right)^{I}{}_{JKL} \bigg[ \left( \theta V^{[K} + \xi A^{[K} \right) e^{L]}  \notag \\
		& & + \ast \left(A^{[K} e^{L]} \right)\bigg],
	\end{eqnarray}
	with
	\begin{equation}
		\left( P^{-1} \right)^{IJ}{}_{KL} = \dfrac{\gamma^{2}}{\gamma^{2} + 1} \left( \delta^{I}_{[K} \delta^{J}_{L]} - \frac{1}{2 \gamma} \epsilon^{IJ}{}_{KL}\right).
	\end{equation}
	Note that the relation between $P^{IJ}{}_{KL}$ and $\left( P^{-1} \right)^{IJ}{}_{KL}$ is given by  
	\begin{equation}
		P^{IJ}{}_{MN} \left( P^{-1} \right)^{MN}{}_{KL} =  \delta^{I}_{[K} \delta^{J}_{L]}.
	\end{equation}
	
	Next, we substitute~\eqref{w_split} into the equations of motion~\eqref{eom_e}, \eqref{eom_p}, and \eqref{eom_bp}. The reader must bear in mind that $F^{I}{}_{J} = \mathcal{R}^{I}{}_{J} + \mathfrak{D} C^{I}{}_{J} + C^{I}{}_{K} \wedge C^{K}{}_{J}$, where $\mathcal{R}^{I}{}_{J} := d \Omega^{I}{}_{J} + \Omega^{I}{}_{K} \wedge \Omega^{K}{}_{J}$ is the curvature of $\Omega^I{}_J$ and $\mathfrak{D}$ is the covariant derivative with respect to $\Omega^{I}{}_{J}$, and that the Bianchi identity for $\omega^I{}_J$ reads $D T^I = F^I{}_J \wedge e^{J}$ while for $\Omega^I{}_J$ it reads $0={\mathcal R}^I{}_J \wedge e^J$. Then, the equations of motion~\eqref{eom_e}, \eqref{eom_p}, and \eqref{eom_bp} become, respectively,
	\begin{subequations}
		\begin{eqnarray}
			&& 2 \kappa  \Big[ e^{J} \wedge \ast \mathcal{R}_{IJ} - \Lambda \star e_{I} + {\ast P}_{IJ}{}^{KL} e^{J} \wedge \big( \mathfrak{D} C_{KL}  \notag \\
			&& + C_{KM} \wedge C^{M}{}_{L} \big) \Big] + \dfrac{1}{2} \bigg[ \bar{\psi} \gamma^{J} E \mathfrak{D} \psi  -  \overline{\mathfrak{D} \psi} \gamma^{J} E^{\dagger} \psi  \notag \\ 
			&& + \dfrac{1}{2} C_{KL} \bar{\psi} \left( \gamma^{J} \sigma^{KL} E + \sigma^{KL} \gamma^{J} E^{\dagger} \right) \psi \bigg]  \wedge \ast \left(e_{I} \wedge e_{J} \right) \notag \\
			\label{eom_e2}
			&& -  m \bar{\psi} \psi \star e_{I} =0, 
		\end{eqnarray}
		\begin{eqnarray}
			&& -  \left( \overline{\mathfrak{D} \psi} - \dfrac{1}{2} C_{JK} \sigma^{JK} \psi \right) \gamma^{I} \wedge \star e_{I}  + \dfrac{1}{2} \bar{\psi} \gamma^{I} E C^{J}{}_{I} \wedge \star e_{J}  \notag \\
			&& -  m \bar{\psi} \rho =0, \label{HolstCD2}\\
			&&  \gamma^{I} \left( \mathfrak{D} \psi + \dfrac{1}{2} C_{JK} \sigma^{JK} \psi \right) \wedge \star e_{I}  - \dfrac{1}{2} \gamma^{I} E^{\dagger}\psi C^{J}{}_{I} \wedge \star e_{J}  \notag \\
			&& - m \psi \rho =0, \label{HolstCD}
		\end{eqnarray}
	\end{subequations}
	with
	\begin{subequations}
		\begin{eqnarray}
			\mathfrak{D} C^{I}{}_{J} &:= & d C^{I}{}_{J} + \Omega^{I}{}_{K} \wedge  C^{K}{}_{J} - \Omega^{K}{}_{J} \wedge C^{I}{}_{K} ,\\
			\mathfrak{D} \psi &:= & d \psi + \dfrac{1}{2} \Omega_{IJ} \sigma^{IJ} \psi,\\
			\overline{\mathfrak{D} \psi} &:= & d \bar{\psi} - \dfrac{1}{2} \Omega_{IJ} \bar{\psi} \sigma^{IJ}.
		\end{eqnarray}
	\end{subequations}
	From~\eqref{S}, we conclude that if gravity were turned off, i.e., if gravity were nondynamical, then the second and third terms of~\eqref{HolstCD} that involve the contorsion $C_{IJ}$ would not appear and~\eqref{HolstCD} would become Dirac's equation for a fermion field propagating on a curved (fixed) background (see also \cite{Kazmierczak0903}). This means that the action principle~\eqref{S} is a suitable generalization to the case when gravity is dynamical.
	
	Note also that~\eqref{eom_e2} implies
	\begin{equation}
		\mathfrak{R}_{IJ} - \dfrac{1}{2} \mathfrak{R} \eta_{IJ}  + \Lambda \eta_{IJ} = \dfrac{1}{2\kappa} T_{IJ},
	\end{equation}
	where $\mathfrak{R}_{IJ}$ is the Ricci tensor, $\mathfrak{R}$ is the scalar curvature\footnote{The Ricci tensor and the scalar curvature are defined by $\mathfrak{R}_{IJ}:= \mathfrak{R}_{KI}{}^{K}{}_{J}$ and $\mathfrak{R}:= \mathfrak{R}^{I}{}_{I}$, respectively, with $\mathcal{R}^{I}{}_{J} = (1/2) \mathfrak{R}_{KL}{}^{I}{}_{J} e^{K} \wedge e^{L}$.}, and 
	\begin{eqnarray}
		T_{IJ} & := & 2 \kappa \Big[  2  P ^{K}{}_{JLM} \left(  \mathfrak{D}_{[I} C_{K]}{}^{LM} + C_{[I}{}^{L}{}_{|N|} C_{K]}{}^{NM}  \right) \notag \\
		& &  + \eta_{I J} P^{KL}{}_{MN} \left( \mathfrak{D}_{K} C_{L}{}^{MN} + C_{K}{}^{M}{}_{P} C_{L}{}^{PN} \right) \Big] \notag \\
		& & - \dfrac{1}{2} \left( \bar{\psi} \gamma_{J} E \mathfrak{D}_{I} \psi - \overline{\mathfrak{D}_{I} \psi} \gamma_{J} E^{\dagger} \psi \right) \notag \\
		& & - \dfrac{1}{4} C_{IKL} \bar{\psi} \big( \gamma_{J} \sigma^{KL} E + \sigma^{K L} \gamma_{J} E^{\dagger} \big) \psi   \notag \\
		& & + \dfrac{1}{2} \eta_{IJ} \Big[ \bar{\psi} \gamma ^{K} E \mathfrak{D}_{K} \psi  - \overline{\mathfrak{D}_{K} \psi} \gamma^{K} E^{\dagger} \psi  \notag  \\
		& &  + \dfrac{1}{2} C_{KLM} \bar{\psi} \big( \gamma^{K} \sigma^{LM} E + \sigma^{LM} \gamma^{K} E^{\dagger} \big) \psi \notag \\
		& & - 2  m \bar{\psi} \psi \Big]  ,
	\end{eqnarray}
	with
	\begin{subequations}
		\begin{eqnarray}
			\mathfrak{D}_{K} C_{L}{}^{I}{}_{J} &:= & \partial_{K} C_{L}{}^{I}{}_{J} + \Omega_{K}{}^{I}{}_{M} C_{L}{}^{M}{}_{J}  \notag \\
			& &- \Omega_{K}{}^{M}{}_{J} C_{L}{}^{I}{}_{M} - \Omega_{K}{}^{M}{}_{L} C_{M}{}^{I}{}_{J}, \\
			\mathfrak{D}_{I} \psi &:= & \partial_{I} \psi + \dfrac{1}{2} \Omega_{IJK} \sigma^{JK} \psi,\\
			\overline{\mathfrak{D}_{I} \psi} &:= & \partial_{I} \bar{\psi} - \dfrac{1}{2} \Omega_{IJK} \bar{\psi} \sigma^{JK}.
		\end{eqnarray}
	\end{subequations}

	The previous definitions arise from expressing $d\psi = e^{I} \partial_{I} \psi $, $\Omega^{I}{}_{J} = \Omega_{K}{}^{I}{}_{J} e^{K}$, $C^{I}{}_{J} = C_{K}{}^{I}{}_{J} e^{K}$, and $\mathfrak{D}C^{I}{}_{J} = \mathfrak{D}_{K} C_{L}{}^{I}{}_{J} e^{K} \wedge e^{L}$.
	
	\subsection{Second-order action}\label{EA}
	The connection $\omega^I{}_J$ is an auxiliary field of the first-order action~\eqref{S} because, from its own equation of motion~\eqref{eom_w}, we can solve for $\omega^I{}_J$ in terms of the other fields as was already shown in~\eqref{w_split}. Therefore, by integrating out the connection $\omega^I{}_J$ in~\eqref{S} by using the solution~\eqref{w_split}, we get an equivalent second-order action principle
	\begin{eqnarray}
		\label{S_eff}
		S_{\mbox{eff}} [e, \psi, \bar{\psi}] & = &  \kappa \int_{M}   \left( \mathfrak{R} - 2 \Lambda \right) \rho +  \int_{M} \bigg[ \dfrac{1}{2} \big( \bar{\psi} \gamma^{I} \mathfrak{D} \psi \notag \\
		& &  - \overline{\mathfrak{D} \psi} \gamma^{I} \psi \big) \wedge \star e_{I}  - m \bar{\psi} \psi \rho \bigg] + S_{\mbox{int}}[e,\bar{\psi}, \psi]\notag \\
		& & - \dfrac{1}{4} \int_{\partial M} \left( \xi A^{I} + \theta V^{I} \right) \star e_{I},
	\end{eqnarray}
	where the interaction action $S_{\mbox{int}}$ is given by
	\begin{eqnarray}
		\label{S_int}
		S_{\rm{int}}[e,\bar{\psi}, \psi] & := & - \dfrac{3 \gamma^{2}}{32 \kappa \left( \gamma^{2} + 1 \right)} \int_{M} \bigg[  - 2 \theta \left(  \xi + \frac{1}{\gamma}  \right) V_{I} A^{I} \notag \\
		& & - \theta^{2} V_{I} V^{I}  +\left(1 - \xi^{2}  - 2 \frac{\xi}{\gamma} \right) A_{I} A^{I} \bigg] \rho. \notag \\
	\end{eqnarray}
	Thus, the second-order action~\eqref{S_eff}, and hence the first-order action~\eqref{S}, is in the generic case different from the Einstein-Dirac theory~\cite{Weyl_1929} (see also~\cite{Weyl_1950}). In the case of the minimal coupling of fermions to gravity, defined by $\theta=0= \xi$, the boundary term in~\eqref{S_eff} vanishes and $S_{\rm{int}}$ only carries the axial-axial interaction modulated by the Immirzi parameter~\cite{Perez0602}.  
	
	\section{Manifestly Lorentz-invariant Hamiltonian formulation}\label{Sec_Ham_form}
	The canonical analysis of the action~\eqref{S} will be obtained by following the same approach applied to the Holst action in Ref.~\cite{Montesinos2004a}. We remind the reader that the idea behind such a canonical analysis is very simple: in such a paper we performed the canonical analysis of the Holst action using a suitable parametrization of the orthonormal frame $e^I$ and the connection $\omega^I{}_J$, which allows us to get straightforwardly its Hamiltonian formulation (involving first-class constraints only) after integrating out the auxiliary fields involved. Such a procedure has the advantage of reducing additionally the presymplectic structure to a canonical one from the very beginning. Once the time gauge is imposed on such a formulation, Barbero's formulation~\cite{Barbero9505} for general relativity arises immediately. In this section, we will show that the approach of Ref.~\cite{Montesinos2004a} can also be applied to the action~\eqref{S} with the corresponding handling of the fermionic contribution. 
	
	Adapted to the spacetime foliation, the orthonormal frame $e^I$ and the connection $\omega^I{}_J$ can be written as $e^I= e_t{}^I dt + e_a{}^I d x^a$ and $\omega^I{}_J = \omega_t{}^I{}_J dt + \omega_a{}^I{}_J d x^a$, respectively. We parametrize the 16 components of the tetrad $e_{\mu}{}^{I}$ in terms of the lapse function $N$, the shift $N^a$, and $\tilde{\Pi}^{aI}$ that is a tensor density of weight $1$ as~\cite{Montesinos2004a}
	\begin{subequations}
		\begin{eqnarray}
			\label{et_rp}
			e_{t}{}^{I} & = & N n^{I} + N^{a} h^{1/4} \uac{\uac{h}}_{ab} \tilde{\Pi}^{bI}, \\
			\label{ea_rp}
			e_{a}{}^{I} & = & h^{1/4} \uac{\uac{h}}_{ab} \tilde{\Pi}^{bI},
		\end{eqnarray} 
	\end{subequations}
	with $n_{I}$ given by 
	\begin{equation} \label{nI}
		n_{I} := \frac{1}{6 \sqrt{h}} \epsilon_{IJKL} \uac{\eta}_{abc} \tilde{\Pi}^{aJ} \tilde{\Pi}^{bK} \tilde{\Pi}^{cL},
	\end{equation}
	where $\uac{\uac{h}}_{ab}$ is the inverse of $\tilde{\tilde{h}}^{ab} = \tilde{\Pi}^{aI} \tilde{\Pi}^{b}{}_{I}$ and $h := \det (\tilde{\tilde{h}}^{ab})$ is a tensor density of weight $4$. Note that the following properties hold: $n_{I} n^{I} = -1 $ and $n_{I} \tilde{\Pi}^{aI} = 0$. The map $(N, N^a, \tilde{\Pi}^{aI}) \mapsto (e_{\mu}{}^I)$ given by~\eqref{et_rp} and~\eqref{ea_rp} is invertible and the inverse map can be found in the Appendix.
	
	Using the $3+1$ decomposition of $e^I$ and $\omega^I{}_J$ together with~\eqref{et_rp} and~\eqref{ea_rp}, the action~\eqref{S} acquires the form
	\begin{eqnarray}
		\label{S_3p1}
		S & = & \int_{\mathbb{R} \times \Sigma} dtd^{3}x \Big[  - 2 \kappa \tilde{\Pi}^{aI} n^{J} \partial_{t} \ovg{\omega}_{aIJ} + \dfrac{1}{2} h^{1/4} n_{I} \bar{\psi} \gamma^{I} E \dot{\psi} \notag \\
		& & - \dfrac{1}{2} h^{1/4} n_{I} \dot{\bar{\psi}} \gamma^{I} E^{\dagger} \psi + \omega_{tIJ} \tilde{\mathcal{G}}^{IJ} - N^{a} \tilde{\mathcal{V}}_{a} - \uac{N} \tilde{\tilde{\mathcal{S}}} \Big],
	\end{eqnarray}
	where $dtd^{3}x := dt \wedge dx^{1} \wedge dx^{2} \wedge dx^{3}$, the dot over the corresponding field denotes $\partial_t$, $\uac{N} := h^{-1/4} N$, and
	\begin{subequations}\label{const_3p1}
		\begin{eqnarray}
			\tilde{\mathcal{G}}^{IJ} & := & - 2 \kappa P^{IJ}{}_{KL} \Big[ \partial_{a} \left( \tilde{\Pi}^{aK} n^{L} \right) + 2 \omega_{a}{}^{K}{}_{M} \tilde{\Pi}^{a[M}n^{L]} \Big] \notag \\
			& & + \dfrac{1}{4} h^{1/4} n_{K} \bar{\psi} \left( \gamma^{K} \sigma^{IJ} E + \sigma^{IJ} \gamma^{K}  E^{\dagger} \right) \psi, \\
			\tilde{\mathcal{V}}_{a} & := & 2 \kappa \Big[ -\tilde{\Pi}^{bI} n^{J} \partial_{a} \ovg{\omega}_{bIJ} + \partial_{b} \Big( \ovg{\omega}_{aIJ} \tilde{\Pi}^{bI} n^{J} \Big) \Big] \notag \\
			& & + \frac{1}{2} h^{1/4} n_{I}  \left( \bar{\psi} \gamma^{I} E \partial_{a} \psi - \partial_{a} \bar{\psi} \gamma^{I} E^{\dagger} \psi \right) \notag \\
			& & + \omega_{aIJ}\tilde{\mathcal{G}}^{IJ}, \\
			\tilde{\tilde{\mathcal{S}}} & := & \kappa \tilde{\Pi}^{aI} \tilde{\Pi}^{bJ} \ovg{F}_{abIJ} + \frac{1}{2} h^{1/4} \tilde{\Pi}^{aI} \Big( \bar{\psi} \gamma_{I} E D_{a} \psi \notag \\
			& &- \overline{D_{a} \psi} \gamma_{I} E^{\dagger} \psi \Big) - h^{1/2} \left( 2 \kappa \Lambda +  m   \bar{\psi} \psi \right),
		\end{eqnarray}
	\end{subequations}
	with
	\begin{eqnarray}
		\label{D_def}
		D_{a} \psi & := & \partial_{a} \psi + \dfrac{1}{2} \omega_{aIJ} \sigma^{IJ} \psi, \\
		\overline{D_{a} \psi} & := & \partial_{a} \bar{\psi} - \dfrac{1}{2} \omega_{aIJ} \bar{\psi} \sigma^{IJ}, \\
		F_{ab}{}^{I}{}_{J} & = & \partial_{a} \omega_{b}{}^{I}{}_{J} - \partial_{b} \omega_{a}{}^{I}{}_{J} + \omega_{a}{}^{I}{}_{K} \omega_{b}{}^{K}{}_{J} \notag \\
		& &  - \omega_{b}{}^{I}{}_{K} \omega_{a}{}^{K}{}_{J}.
	\end{eqnarray}
	
	Following Refs.~\cite{Montesinos1801,Montesinos2004a}, we write the gravitational part of the presymplectic structure of~\eqref{S_3p1} as
	\begin{equation}
		- 2 \kappa \tilde{\Pi}^{aI} n^{J} \partial_{t} \ovg{\omega}_{aIJ} =  2 \kappa \tilde{\Pi}^{aI} \partial_{t} \left( W_{a}{}^{b}{}_{IJK} \ovg{\omega}_{b}{}^{JK} \right),
	\end{equation}
	from which we observe that $\tilde{\Pi}^{aI}$ is the momentum canonically conjugate to the configuration variable $C_{aI}$, which is defined by~\cite{Montesinos1801, Montesinos2004a}
	\begin{equation}
		\label{C_def}
		C_{aI} := W_{a}{}^{b}{}_{IJK} \ovg{\omega}_{b}{}^{JK},
	\end{equation}
	where
	\begin{equation}
		\label{W_def}
		W_{a}{}^{b}{}_{IJK} := - \delta^{b}_{a} \eta_{I[J}n_{K]} - n_{I} \uac{\uac{h}}_{ac} \tilde{\Pi}^{c}{}_{[J}  \tilde{\Pi}^{b}{}_{K]}.
	\end{equation}
	Once the reduction of the gravitational part of the presymplectic structure has been achieved, it remains to parametrize $\ovg{\omega}_{aIJ}$ in terms of the configuration variable $C_{aI}$ and six additional fields $\uac{\lambda}_{ab}$ ($= \uac{\lambda}_{ba}$). This is done by solving~\eqref{C_def}, which gives~\cite{Montesinos2004a}
	\begin{equation}\label{w}
		\ovg{\omega}_{aIJ} = M_{a}{}^{b}{}_{IJK} C_{b}{}^{K} + \tilde{N}^{b}{}_{IJ} \uac{\lambda}_{ab},
	\end{equation}
	where
	\begin{eqnarray}\label{M}
		M_{a}{}^{b}{}_{IJK} & := & - \delta^{b}_{a} n_{[I} \eta_{J]K} + \uac{\uac{h}}_{ac} \tilde{\Pi}^{b}{}_{[I}   \tilde{\Pi}^{c}{}_{J]} n_{K} \hspace{50pt} \quad \quad \notag \\
		& & + \delta^{b}_{a} P_{IJKL} n^{L} + \frac{1}{2 \gamma} \epsilon_{IJLM} \uac{\uac{h}}_{ac} \tilde{\Pi}^{c}{}_{K} \tilde{\Pi}^{bL} n^{M},\\
		\label{N}
		\tilde{N}^{a}{}_{IJ} & := & \epsilon_{IJKL} \tilde{\Pi}^{aK} n^{L}.
	\end{eqnarray}
	See the Appendix for the inverse map of~\eqref{w} and additional algebraic properties of the objects involved in these maps. 
	
	The following step is to substitute~\eqref{w} into the action~\eqref{S_3p1}, obtaining
	\begin{eqnarray}
		\label{S_wG}
		S & = & \int_{\mathbb{R} \times \Sigma} dtd^{3}x \bigg[  2 \kappa \tilde{\Pi}^{aI} \dot{C}_{aI} + \dfrac{1}{2} h^{1/4} n_{I} \Big( \bar{\psi} \gamma^{I} E \dot{\psi} \notag \\
		& &  -  \dot{\bar{\psi}} \gamma^{I} E^{\dagger} \psi \Big) + \omega_{tIJ} \tilde{\mathcal{G}}^{IJ} - N^{a} \tilde{\mathcal{V}}_{a} - \uac{N} \tilde{\tilde{\mathcal{S}}} \bigg],
	\end{eqnarray}
	with 
	\begin{widetext}
		\begin{subequations} 
			\label{const_wG}
			\begin{eqnarray}
				\tilde{\mathcal{G}}^{IJ} & = & 2 \kappa \Big( \tilde{\Pi}^{a[I} C_{a}{}^{J]} + 2 P^{IJ}{}_{KL} \tilde{\Pi}^{a[K} n^{M]} \Gamma_{a}{}^{L}{}_{M} \Big) 
				+ \dfrac{1}{4} h^{1/4} n_{K} \bar{\psi} \left( \gamma^{K} \sigma^{IJ} E + \sigma^{IJ} \gamma^{K}  E^{\dagger} \right) \psi, \\
				\tilde{\mathcal{V}}_{a} & = & 2 \kappa \Big( 2 \tilde{\Pi}^{bI} \partial_{[a} C_{b]I} - C_{aI} \partial_{b} \tilde{\Pi}^{bI} \Big) + \frac{1}{2} h^{1/4} n_{I}  \left( \bar{\psi} \gamma^{I} E  \partial_{a} \psi - \partial_{a} \bar{\psi} \gamma^{I} E^{\dagger} \psi \right) \notag \\
				& & + \left(P^{-1}\right)_{IJKL} \Big( M_{a}{}^{bKLM} C_{bM} + \tilde{N}^{bKL} \uac{\lambda}_{ab}\Big) \tilde{\mathcal{G}}^{IJ},  \\
				\tilde{\tilde{\mathcal{S}}} & = & \kappa \tilde{\Pi}^{aI} \tilde{\Pi}^{bJ} R_{abIJ} + 2 \kappa \tilde{\Pi}^{a[I} \tilde{\Pi}^{|b|J]} \bigg[ C_{aI} C_{bJ} + 2 C_{aI} \ovg{\Gamma}_{bJK} n^{K} + \frac{1}{\gamma^{2}} q^{KL} \Gamma_{aIK} \Gamma_{bJL} + \bigg( \Gamma_{aIK} + \dfrac{2}{\gamma} \ast \Gamma_{aIK} \bigg)  \notag \\
				& & \times \Gamma_{bJL} n^{K} n^{L} \bigg] + \dfrac{1}{2} h^{1/4} \tilde{\Pi}^{aI} \left( \bar{\psi} \gamma_{I} \nabla_{a} \psi - \overline{\nabla_{a} \psi} \gamma_{I} \psi \right) + \dfrac{1}{2} h^{1/4}  \Big( C_{aI} + \ovg{\Gamma}_{aIK}n^{K} \Big) n_{J} \tilde{\Pi}^{a}{}_{L} \bar{\psi} \Big( \gamma^{L} \sigma^{IJ} E + \sigma^{IJ} \gamma^{L} E^{\dagger} \Big) \psi \notag \\
				& & - \dfrac{3}{32 \kappa} h^{1/2} q_{IJ} \bigg\lbrace - A^{I} A^{J}  + \dfrac{\gamma^{2}}{\gamma^{2} + 1 } \bigg[ - 2 \theta \left(  \xi + \frac{1}{\gamma} \right) A^{I} V^{J}  - \theta^{2} V^{I} V^{J} + \left( 1 - \xi^{2} - 2\frac{\xi}{\gamma}  \right) A^{I} A^{J}\bigg] \bigg\rbrace  \notag \\
				& & - \bigg( \uac{\lambda}_{ab} - \uac{U}_{ab}{}^{fIJ} \ovg{\Gamma}_{fIJ} \bigg) \bigg[ \dfrac{\kappa \gamma^{2}}{\gamma^{2} + 1} G^{abcd} \bigg( \uac{\lambda}_{cd} - \uac{U}_{cd}{}^{eKL} \ovg{\Gamma}_{eKL}\bigg)  - \dfrac{1}{2} h^{1/4}  \ast \left(P^{-1}\right)_{KLMN} \tilde{\Pi}^{a}{}_{P} \tilde{\Pi}^{bM} n^{N}  \notag \\
				& & \times \bar{\psi}\bigg( \gamma^{P} \sigma^{KL} E + \sigma^{KL} \gamma^{P} E^{\dagger} \bigg) \psi \bigg] - \dfrac{1}{4 \kappa} \tilde{\mathcal{G}}^{IJ} \bigg\lbrace \tilde{\mathcal{G}}_{IJ} - \left( P^{-1} \right)_{IJKL} \tilde{\mathcal{G}}^{KL} - 2 n_{I} n^{K} \tilde{\mathcal{G}}_{JK}  - h^{1/4} \bigg[ \left( P^{-1} \right)_{IJKL} n^{K}  \big( \theta V^{L} \notag \\
				\label{H_wG}
				& & +\xi A^{L} \big) + \ast \left( P^{-1} \right)_{IJKL} n^{K} A^{L} - \dfrac{1}{2} \epsilon_{IJKL}  n^{K} A^{L}  \bigg] \bigg\rbrace + 2 \tilde{\Pi}^{aI} n^{J} \nabla_{a} \tilde{\mathcal{G}}_{IJ}  - h^{1/2} \left( 2 \kappa \Lambda + m \bar{\psi} \psi  \right). 
			\end{eqnarray}
		\end{subequations}
	\end{widetext}
	In the previous expressions we have introduced the covariant derivative $\nabla_{a}$ defined by
	\begin{eqnarray} 
		\label{CD}
		\nabla_{a} \tilde{\Pi}^{bI} &:=& \partial_{a} \tilde{\Pi}^{bI} + \Gamma^{b}{}_{ca} \tilde{\Pi}^{cI} - \Gamma^{c}{}_{ca} \tilde{\Pi}^{bI} + \Gamma_{a}{}^{I}{}_{J} \tilde{\Pi}^{bJ} \notag\\
		&=&0,
	\end{eqnarray}
	and $\Gamma^{a}{}_{bc} = \Gamma^{a}{}_{cb}$ and $\Gamma_{aIJ} = -\Gamma_{aJI}$. See Ref.~\cite{Montesinos2004a} for the explicit solutions of $\Gamma^{a}{}_{bc}$ and $\Gamma_{aIJ}$. The curvature of $\Gamma_{aIJ}$ is given by $R_{ab}{}^{I}{}_{J} := \partial_{a} \Gamma_{b}{}^{I}{}_{J} - \partial_{b} \Gamma_{a}{}^{I}{}_{J} + \Gamma_{a}{}^{I}{}_{K} \Gamma_{b}{}^{K}{}_{J} - \Gamma_{b}{}^{I}{}_{K} \Gamma_{a}{}^{K}{}_{J}$, and we have also introduced the following expressions
	\begin{eqnarray}
		\label{n_psi}
		\nabla_{a} \psi & := & \partial_{a} \psi + \dfrac{1}{2} \Gamma_{aIJ} \sigma^{IJ} \psi, \\
		\label{n_bpsi}
		\overline{\nabla_{a} \psi} & := & \partial_{a} \bar{\psi} - \dfrac{1}{2} \Gamma_{aIJ} \bar{\psi} \sigma^{IJ},  \\
		\nabla_{a} \tilde{\mathcal{G}}^{IJ} & := & \partial_{a} \tilde{\mathcal{G}}^{IJ} - \Gamma^{b}{}_{ba} \tilde{\mathcal{G}}^{IJ} + \Gamma_{a}{}^{I}{}_{K} \tilde{\mathcal{G}}^{KJ} \notag \\ 
		&&+ \Gamma_{a}{}^{J}{}_{K} \tilde{\mathcal{G}}^{IK},  \\
		q^{I}{}_{J} & := & \delta^{I}_{J} + n^{I} n_{J}, \\
		G^{abcd} & := & \tilde{\tilde{h}}^{ab} \tilde{\tilde{h}}^{cd} - \tilde{\tilde{h}}^{c(a} \tilde{\tilde{h}}^{b)d}, \\
		\label{U}
		\uac{U}_{ab}{}^{cIJ} &:=&  \left(1+\frac{1}{\gamma^2}\right)\ast(P^{-1})^{IJKL}  \delta^{c}{}_{(a} \uac{\uac{h}}_{b)e}\tilde{\Pi}^{e}{}_{K} n_{L},
	\end{eqnarray}
	where $G^{abcd}$ is a tensor density of weight $4$. Note that the reduction map $(\omega_{aIJ}, \tilde{\Pi}^{aI}) \mapsto (C_{aI}, \tilde{\Pi}^{aI})$ given by~\eqref{C_def} leaves the null directions of the presymplectic structure~\eqref{S_3p1} out of the symplectic structure, which are clearly along $\uac{\lambda}_{ab}$.
	
	We now integrate by parts the term containing the covariant derivative of $\tilde{\mathcal{G}}^{IJ}$ in~\eqref{H_wG} and factor out all the terms involving $\tilde{\mathcal{G}}^{IJ}$ in \eqref{S_wG}, which requires to redefine the Lagrange multiplier $\omega_{tIJ}$ as 
	\begin{eqnarray}\label{l}
		\omega_{tIJ} & =: & - \lambda_{IJ} + N^{a} \left(P^{-1}\right)_{IJKL} \Big( M_{a}{}^{bKLM} C_{bM}  \notag \\
		& & + \tilde{N}^{bKL} \uac{\lambda}_{ab}\Big) - \dfrac{1}{4 \kappa} \uac{N} \bigg\lbrace \tilde{\mathcal{G}}_{IJ} - \left( P^{-1} \right)_{IJKL} \tilde{\mathcal{G}}^{KL} \notag \\
		& & - 2 n^{K} n_{[I}  \tilde{\mathcal{G}}_{J]K}   - h^{1/4} \bigg[   \ast \left( P^{-1} \right)_{IJKL} n^{K} A^{L}  \notag \\
		& & + \left( P^{-1} \right)_{IJKL} n^{K}  \big( \theta V^{L} + \xi A^{L} \big)   \notag \\
		& &   - \dfrac{1}{2} \epsilon_{IJKL}  n^{K} A^{L}  \bigg] \bigg\rbrace - 2 \tilde{\Pi}^{a}{}_{[I} n_{J]} \nabla_{a} \uac{N}. 
	\end{eqnarray}
	\newpage
	\noindent
	Note that the map $(C_{aI}, \uac{\lambda}_{ab}, \lambda_{IJ}) \mapsto (\omega_{\mu}{}^{I}{}_{J})$ given by~\eqref{w} and~\eqref{l} gives the parametrization of $\omega_{\mu}{}^{I}{}_{J}$ as a function of 12 configuration variables $C_{aI}$, 6 fields $\uac{\lambda}_{ab}$, and 6 Lagrange multipliers $\lambda_{IJ}$. Therefore, the action can be equivalently written as
	\begin{eqnarray}
		\label{S_wl}
		S & = & \int_{\mathbb{R} \times \Sigma} dtd^{3}x \bigg[  2 \kappa \tilde{\Pi}^{aI} \dot{C}_{aI} + \dfrac{1}{2} h^{1/4} n_{I} \Big( \bar{\psi} \gamma^{I} E \dot{\psi} \notag \\
		& &  -  \dot{\bar{\psi}} \gamma^{I} E^{\dagger} \psi \Big) - \lambda_{IJ} \tilde{\mathcal{G}}^{IJ} - 2 N^{a} \tilde{\mathcal{D}}_{a} - \uac{N} \tilde{\tilde{\mathcal{Z}}} \bigg],
	\end{eqnarray}
	where $\tilde{\mathcal{G}}^{IJ}$, $\tilde{\mathcal{D}}_{a}$, and $\tilde{\tilde{\mathcal{Z}}}$ are given, respectively, by 
	\begin{subequations}
		\label{const_woG}
		\begin{eqnarray}
			\tilde{\mathcal{G}}^{IJ} & = & 2 \kappa \Big( \tilde{\Pi}^{a[I} C_{a}{}^{J]} + 2 P^{IJ}{}_{KL} \tilde{\Pi}^{a[K} n^{M]} \Gamma_{a}{}^{L}{}_{M} \Big) \notag \\
				&& + \dfrac{1}{4} h^{1/4} n_{K} \bar{\psi} \left( \gamma^{K} \sigma^{IJ} E + \sigma^{IJ} \gamma^{K}  E^{\dagger} \right) \psi, \\
			\tilde{\mathcal{D}}_{a} & := & \kappa \Big( 2 \tilde{\Pi}^{bI} \partial_{[a} C_{b]I} - C_{aI} \partial_{b} \tilde{\Pi}^{bI} \Big) \notag \\
				& &+ \frac{1}{4} h^{1/4} n_{I}  \left( \bar{\psi} \gamma^{I} E  \partial_{a} \psi - \partial_{a} \bar{\psi} \gamma^{I} E^{\dagger} \psi \right), 
		\end{eqnarray}	
		\begin{widetext}	
		\begin{eqnarray}
				\tilde{\tilde{\mathcal{Z}}} & := & \kappa \tilde{\Pi}^{aI} \tilde{\Pi}^{bJ} R_{abIJ} + 2 \kappa \tilde{\Pi}^{a[I} \tilde{\Pi}^{|b|J]} \bigg[ C_{aI} C_{bJ} + 2 C_{aI} \ovg{\Gamma}_{bJK} n^{K} + \frac{1}{\gamma^{2}} q^{KL} \Gamma_{aIK} \Gamma_{bJL} + \bigg( \Gamma_{aIK} + \dfrac{2}{\gamma} \ast \Gamma_{aIK} \bigg)  \notag \\
				& & \times \Gamma_{bJL} n^{K} n^{L} \bigg] + \dfrac{1}{2} h^{1/4} \tilde{\Pi}^{aI} \left( \bar{\psi} \gamma_{I} \nabla_{a} \psi - \overline{\nabla_{a} \psi} \gamma_{I} \psi \right) + \dfrac{1}{2} h^{1/4}  \Big( C_{aI} + \ovg{\Gamma}_{aIK}n^{K} \Big) n_{J} \tilde{\Pi}^{a}{}_{L} \bar{\psi} \Big( \gamma^{L} \sigma^{IJ} E + \sigma^{IJ} \gamma^{L} E^{\dagger} \Big) \psi \notag \\
				& & - \dfrac{3}{32 \kappa} h^{1/2} q_{IJ} \bigg\lbrace - A^{I} A^{J}  + \dfrac{\gamma^{2}}{\gamma^{2} + 1 } \bigg[ - 2 \theta \left(  \xi + \frac{1}{\gamma} \right) A^{I} V^{J}  - \theta^{2} V^{I} V^{J} + \left( 1 - \xi^{2} - 2 \frac{\xi}{\gamma}  \right) A^{I} A^{J}\bigg] \bigg\rbrace  \notag \\
				& & - \bigg( \uac{\lambda}_{ab} - \uac{U}_{ab}{}^{fIJ} \ovg{\Gamma}_{fIJ} \bigg) \bigg[ \dfrac{\kappa \gamma^{2}}{\gamma^{2} + 1} G^{abcd} \bigg( \uac{\lambda}_{cd} - \uac{U}_{cd}{}^{eKL} \ovg{\Gamma}_{eKL}\bigg)  - \dfrac{1}{2} h^{1/4}  \ast \left(P^{-1}\right)_{KLMN} \tilde{\Pi}^{a}{}_{P} \tilde{\Pi}^{bM} n^{N}  \notag \\
				& & \times \bar{\psi}\bigg( \gamma^{P} \sigma^{KL} E + \sigma^{KL} \gamma^{P} E^{\dagger} \bigg) \psi \bigg]  - h^{1/2} \left( 2 \kappa \Lambda + m \bar{\psi} \psi  \right). 
			\end{eqnarray}
		\end{widetext}	
		\end{subequations}
	
	To cast the action into the desired Hamiltonian form, we must deal with the variables $\uac{\lambda}_{ab}$, which appear in the action \eqref{S_wl} in a quadratic fashion. Although we could apply the cumbersome and lengthy Dirac's method~\cite{dirac1964lectures} and define the momentum canonically conjugate to $\uac{\lambda}_{ab}$, this way of proceeding would lead us to the introduction of second-class constraints, which would still have to be explicitly solved or handled with the Dirac bracket. Once we solve such second-class constraints we would arrive to what is obtained by simply integrating out the auxiliary field $\uac{\lambda}_{ab}$. This is why it makes no sense to follow Dirac's approach to handle $\uac{\lambda}_{ab}$. Therefore, following the approach of Refs.~\cite{Montesinos2004a, Montesinos2001}, we integrate out $\uac{\lambda}_{ab}$ in~\eqref{S_wl}. To do so, we make the variation of~\eqref{S_wl} with respect to $\uac{\lambda}_{ab}$, which leads to its equation of motion
	\begin{eqnarray}\label{eom_l}
		& & \dfrac{ 2 \kappa \gamma^{2}}{\gamma^{2} + 1} \uac{N} G^{abcd} \bigg( \uac{\lambda}_{cd} - \uac{U}_{cd}{}^{eIJ} \ovg{\Gamma}_{eIJ}\bigg)  \notag \\
		& & - \dfrac{1}{2} \uac{N} h^{1/4} \ast \left(P^{-1}\right)_{IJKL} \tilde{\Pi}^{(a}{}_{M} \tilde{\Pi}^{b)K} n^{L} \bar{\psi}\bigg( \gamma^{M} \sigma^{IJ} E \notag \\
		& &	+ \sigma^{IJ} \gamma^{M} E^{\dagger} \bigg) \psi =0.
	\end{eqnarray}
	Therefore, since $\uac{N} \neq 0$, we get
	\begin{eqnarray}\label{sol_l}
		\uac{\lambda}_{ab} & = &  \uac{U}_{ab}{}^{cIJ} \ovg{\Gamma}_{cIJ} + \dfrac{h^{1/4}}{4 \kappa} \left(1 + \dfrac{1}{\gamma^{2}} \right)   \ast \left(P^{-1}\right)_{IJKL} \notag \\
		& & \times \left(G^{-1}\right)_{abcd}  \tilde{\Pi}^{c}{}_{M} \tilde{\Pi}^{dK} n^{L} \bar{\psi}\bigg( \gamma^{M} \sigma^{IJ} E \notag \\
		& &	+ \sigma^{IJ} \gamma^{M} E^{\dagger} \bigg) \psi.
	\end{eqnarray}
	where $\left(G^{-1}\right)_{abcd} = (1/2) (\uac{\uac{h}}_{ab} \uac{\uac{h}}_{cd} - 2 \uac{\uac{h}}_{c(a} \uac{\uac{h}}_{b)d})$ is a tensor density of weight $-4$. Note that $G^{abcd}$ and  $\left(G^{-1}\right)_{abcd}$ satisfy $G^{abcd} \left(G^{-1}\right)_{cdef}= \delta^{a}_{(e} \delta^{b}_{f)}$.
	
	Substituting~\eqref{sol_l} into the action~\eqref{S_wl}, we get
	\begin{eqnarray}
		\label{S_integrated}
		S & = & \int_{\mathbb{R} \times \Sigma} dtd^{3}x \bigg[  2 \kappa \tilde{\Pi}^{aI} \dot{C}_{aI} + \dfrac{1}{2} h^{1/4} n_{I} \Big( \bar{\psi} \gamma^{I} E \dot{\psi} \notag \\
		& &  -  \dot{\bar{\psi}} \gamma^{I} E^{\dagger} \psi \Big) - \lambda_{IJ} \tilde{\mathcal{G}}^{IJ} - 2 N^{a} \tilde{\mathcal{D}}_{a} - \uac{N} \tilde{\tilde{\mathcal{H}}} \bigg],
	\end{eqnarray}
	where the Gauss $\tilde{\mathcal{G}}^{IJ}$, diffeomorphism $\tilde{\mathcal{D}}_{a}$, and Hamiltonian $\tilde{\tilde{\mathcal{H}}}$ constraints are given, respectively, by
	\begin{subequations}
			\label{const_nc}
			\begin{eqnarray}
				\label{const_nc_G}
				\tilde{\mathcal{G}}^{IJ} & = & 2 \kappa \Big( \tilde{\Pi}^{a[I} C_{a}{}^{J]} + 2P^{IJ}{}_{KL} \tilde{\Pi}^{a[K} n^{M]} \Gamma_{a}{}^{L}{}_{M} \Big) \notag \\
				&& + \dfrac{1}{4} h^{1/4} n_{K} \bar{\psi} \left( \gamma^{K} \sigma^{IJ} E + \sigma^{IJ} \gamma^{K}  E^{\dagger} \right) \psi, \\
				\label{const_nc_D}
				\tilde{\mathcal{D}}_{a} & = &  \kappa \Big( 2 \tilde{\Pi}^{bI} \partial_{[a} C_{b]I} - C_{aI} \partial_{b} \tilde{\Pi}^{bI} \Big) \notag \\
				&&+ \frac{1}{4} h^{1/4} n_{I}  \left( \bar{\psi} \gamma^{I} E  \partial_{a} \psi - \partial_{a} \bar{\psi} \gamma^{I} E^{\dagger} \psi \right), 
			\end{eqnarray}
		\begin{widetext}
			\begin{eqnarray}
				\tilde{\tilde{\mathcal{H}}} & := & \kappa \tilde{\Pi}^{aI} \tilde{\Pi}^{bJ} R_{abIJ} + 2 \kappa \tilde{\Pi}^{a[I} \tilde{\Pi}^{|b|J]} \bigg[ C_{aI} C_{bJ} + 2 C_{aI} \ovg{\Gamma}_{bJK} n^{K} + \frac{1}{\gamma^{2}} q^{KL} \Gamma_{aIK} \Gamma_{bJL} + \bigg( \Gamma_{aIK} + \dfrac{2}{\gamma} \ast \Gamma_{aIK} \bigg) \notag \\
				& & \times \Gamma_{bJL} n^{K} n^{L} \bigg]  + \dfrac{1}{2} h^{1/4} \tilde{\Pi}^{aI} \left( \bar{\psi} \gamma_{I} \nabla_{a} \psi - \overline{\nabla_{a} \psi} \gamma_{I} \psi \right) + \dfrac{1}{2} h^{1/4}  \Big( C_{aI} + \ovg{\Gamma}_{aIK}n^{K} \Big) n_{J} \tilde{\Pi}^{a}{}_{L} \bar{\psi} \Big( \gamma^{L} \sigma^{IJ} E + \sigma^{IJ} \gamma^{L} E^{\dagger} \Big) \psi \notag \\
				& & + \dfrac{3}{32 \kappa} h^{1/2} \bigg\lbrace  q_{IJ} A^{I} A^{J}  + n_{I} n_{J} \big( \theta^{2} V^{I} V^{J} + \xi^{2} A^{I} A^{J} + 2 \theta \xi A^{I}  V^{J} \big) - \dfrac{\gamma^{2}}{\gamma^{2} + 1 } \bigg[  - 2 \theta \left(  \xi + \frac{1}{\gamma} \right) A_{I} V^{I}  - \theta^{2} V_{I} V^{I} \notag \\
				\label{const_nc_H}
				& & +  \left( 1 - \xi^{2} - 2 \frac{\xi}{\gamma}  \right) A_{I} A^{I}  \bigg] \bigg\rbrace  - h^{1/2} \left( 2 \kappa \Lambda + m \bar{\psi} \psi  \right).
			\end{eqnarray}	
			\end{widetext}
		\end{subequations}

	The action~\eqref{S_integrated}, together with~\eqref{const_nc_G}-\eqref{const_nc_H}, is a manifestly Lorentz-invariant Hamiltonian formulation of the Holst action with a nonminimally coupled fermion field. It has very appealing properties. It shows that the coupling of the fermion field to general relativity at the Hamiltonian level involves a real noncanonical symplectic structure. Furthermore, the phase-space variables of the gravitational sector $(C_{aI}, \tilde{\Pi}^{aI} )$ are canonical and real, and they have the same form than the ones for general relativity with a cosmological constant~\cite{Montesinos1801, Montesinos2004a}. Moreover, in the symplectic structure there are manifestly Lorentz-covariant nontrivial interacting terms between the fermion field and the gravitational momentum.
	
	It is important to remark again that the current approach is an alternative road to Dirac's classical analysis. It has the advantage that simplifies significantly the Hamiltonian analysis. One of the key aspects of our approach is the fact that the reduction of the presymplectic structure encoded in~\eqref{C_def} suggests to parametrize the spatial part of the connection $\omega_a{}^I{}_J$ in the form given by~\eqref{w}, and once~\eqref{w} is substituted into the action principle, we realize that $\uac{\lambda}_{ab}$ is an auxiliary field that can be simply integrated out. In this way, we arrive straightforwardly at the Hamiltonian formulation with only first-class constraints, which generate the distinctive gauge symmetries of the theory (local Lorentz transformations and spacetime diffeomorphisms).
	
	We can go one step further and rewrite the constraints by using the explicit form of the coupling matrix $E$ and the identity 
	\begin{equation}
		\label{gammas_id}
		\gamma^{I} \gamma^{J} \gamma^{K}  =  \eta^{IJ} \gamma^{K} - \eta^{IK} \gamma^{J} + \eta^{JK} \gamma^{I}  + \mathrm{i} \, \epsilon^{IJKL} \gamma^{5} \gamma_{L},
	\end{equation}
	and we get
	\begin{widetext}
		\begin{subequations} 
			\label{const_nc_AV}
			\begin{eqnarray}
				\label{const_nc_AV_G}
				\tilde{\mathcal{G}}^{IJ} & = & 2 \kappa \Big( \tilde{\Pi}^{a[I} C_{a}{}^{J]} + 2 P^{IJ}{}_{KL} \tilde{\Pi}^{a[K} n^{M]} \Gamma_{a}{}^{L}{}_{M} \Big) 
				+ \dfrac{1}{2} h^{1/4} n^{[I} \left( \theta V^{J]} + \xi A^{J]} \right) + \dfrac{1}{4} h^{1/4} \epsilon^{IJ}{}_{KL} n^{K} A^{L},  \\
				\label{const_nc_AV_D}
				\tilde{\mathcal{D}}_{a} & = &  \kappa \Big( 2 \tilde{\Pi}^{bI} \partial_{[a} C_{b]I} - C_{aI} \partial_{b} \tilde{\Pi}^{bI} \Big) + \frac{1}{4} h^{1/4} n_{I}  \left[  \bar{\psi} \gamma^{I} \partial_{a} \psi - \partial_{a} \bar{\psi} \gamma^{I} \psi + \partial_{a} \left( \theta V^{I} + \xi A^{I} \right) \right], \\
				\tilde{\tilde{\mathcal{H}}} & = & \kappa \tilde{\Pi}^{aI} \tilde{\Pi}^{bJ} R_{abIJ} + 2 \kappa \tilde{\Pi}^{a[I} \tilde{\Pi}^{|b|J]} \bigg[ C_{aI} C_{bJ} + 2 C_{aI} \ovg{\Gamma}_{bJK} n^{K} + \frac{1}{\gamma^{2}} q^{KL} \Gamma_{aIK} \Gamma_{bJL} + \bigg( \Gamma_{aIK} + \dfrac{2}{\gamma} \ast \Gamma_{aIK} \bigg) \notag \\
				& &\times \Gamma_{bJL} n^{K} n^{L} \bigg] + \dfrac{1}{2} h^{1/4} \tilde{\Pi}^{aI} \left( \bar{\psi} \gamma_{I} \nabla_{a} \psi - \overline{\nabla_{a} \psi} \gamma_{I} \psi \right) + \dfrac{1}{2} h^{1/4} \Big( C_{aI} + \ovg{\Gamma}_{aIM}n^{M} \Big) n_{J} \Big[ \tilde{\Pi}^{aI} \left( \theta V^{J} + \xi A^{J} \right)   \notag \\
				& & +  \epsilon^{IJ}{}_{KL} \tilde{\Pi}^{aK} A^{L} \Big]  + \dfrac{3}{32 \kappa} h^{1/2} \bigg\lbrace q_{IJ} A^{I} A^{J}  + n_{I} n_{J} \big( \theta^{2} V^{I} V^{J} + \xi^{2} A^{I} A^{J} + 2 \theta \xi A^{I}  V^{J} \big)   \notag \\
				\label{const_nc_AV_H}
				& & - \dfrac{\gamma^{2}}{\gamma^{2} + 1 } \bigg[- 2 \theta \left(  \xi + \frac{1}{\gamma} \right) V_{I} A^{I} - \theta^{2} V_{I} V^{I} + \left( 1 - \xi^{2} - 2 \frac{\xi}{\gamma} \right) A_{I}A^{I} \bigg] \bigg\rbrace  - h^{1/2} \left( 2 \kappa \Lambda +  m \bar{\psi} \psi  \right).
			\end{eqnarray}
		\end{subequations}
	\end{widetext}
	This formulation explicitly displays the role played by the coupling parameters $\theta$ and $\xi$.

	\section{Hamiltonian formulation involving half-densitized fermion fields}\label{Sec_hf}
	
	The use of half-densitized fermion fields in the canonical theory of gravity with fermion fields has been put forward and championed by Thiemann in both the first-order formalism~\cite{Thiemann_1998_1487} (see also~\cite{Bojowald0809}) and the second-order formalism~\cite{Thiemann_1998_1281}. Motivated by this fact, in this section, we give a manifestly Lorentz-invariant Hamiltonian formulation for general relativity with fermion fields that uses half-densitized fermion fields. 
	
	The idea is to express the Hamiltonian formulation given by the action~\eqref{S_integrated} and the constraints~\eqref{const_nc_AV_G}-\eqref{const_nc_AV_H} in terms of half-densitized fermion fields. Therefore, we define the half-densitized fermion fields
	\begin{subequations}
		\begin{eqnarray}
			\label{hf1}
			\phi &:= & h^{1/8} \psi, \\
			\label{hf2}
			\bar{\phi} &:= & h^{1/8} \bar{\psi}, 
		\end{eqnarray}
	\end{subequations}
	and we write the noncanonical symplectic structure~\eqref{S_integrated} in terms of them as
	\begin{eqnarray}
		&& 2 \kappa \tilde{\Pi}^{aI} \dot{C}_{aI} + \dfrac{1}{2} h^{1/4} n_{I} \Big( \bar{\psi} \gamma^{I} E \dot{\psi}  -  \dot{\bar{\psi}} \gamma^{I} E^{\dagger} \psi \Big) \notag \\
		&=&	2 \kappa \tilde{\Pi}^{aI} \partial_{t} \left[  C_{aI} + \dfrac{1}{16 \kappa} \uac{\uac{h}}_{ab} \tilde{\Pi}^{b}{}_{I} n_{J} \bar{\phi} \gamma^{J} \left( E-  E^{\dagger} \right)  \phi \right] \notag \\
		& & + \dfrac{1}{2} n_{I} \Big( \bar{\phi} \gamma^{I} E \dot{\phi}  -  \dot{\bar{\phi}} \gamma^{I} E^{\dagger} \phi \Big) \nonumber\\
		&& - \dfrac{3}{8} \partial_{t}\left[  n_{I} \bar{\phi} \gamma^{I} \left( E - E^{\dagger} \right) \phi \right].
	\end{eqnarray}
	From this, we realize that it is natural to define a new {\it real} configuration variable $\Phi_{aI}$ for the gravitational sector as
	\begin{equation}
		\label{Phi_def}
		\Phi_{aI} := C_{aI} + \dfrac{1}{16 \kappa} \uac{\uac{h}}_{ab} \tilde{\Pi}^{b}{}_{I} n_{J} \bar{\phi} \gamma^{J} \left( E-  E^{\dagger} \right)  \phi.
	\end{equation}
	Using this definition, the noncanonical symplectic structure given in~\eqref{S_integrated} acquires the form 
	\begin{eqnarray}
		\label{ktP}
		&& 2 \kappa \tilde{\Pi}^{aI} \dot{C}_{aI} + \dfrac{1}{2} h^{1/4} n_{I} \Big( \bar{\psi} \gamma^{I} E \dot{\psi}  -  \dot{\bar{\psi}} \gamma^{I} E^{\dagger} \psi \Big) \notag \\
		&=& 2 \kappa \tilde{\Pi}^{aI} \dot{\Phi}_{aI} + \dfrac{1}{2} n_{I} \Big( \bar{\phi} \gamma^{I} E \dot{\phi}  -  \dot{\bar{\phi}} \gamma^{I} E^{\dagger} \phi \Big) \nonumber\\
		&& - \dfrac{3}{8} \partial_{t}\left[  n_{I} \bar{\phi} \gamma^{I} \left( E - E^{\dagger} \right) \phi \right].
	\end{eqnarray}
	Note that the last line of~\eqref{ktP} is real\footnote{In terms of the real fermion currents, the last line of~\eqref{ktP} is rewritten as $$- \dfrac{3}{8} \partial_{t}\left[  n_{I} \bar{\phi} \gamma^{I} \left( E - E^{\dagger} \right) \phi \right] = - \dfrac{3}{4} \partial_{t}\left[  h^{1/4} n_{I} \left( \theta V^{I} + \xi A^{I} \right) \right]. $$} and if we neglect it, the new symplectic structure is still real (this type of terms have also been dropped in previous canonical approaches to the coupling of fermion fields to gravity~\cite{Thiemann_1998_1487,Bojowald0809} and even for gravity alone~\cite{Peldan9400}).
	
	Therefore, using half-densitized fermion fields and the new configuration variable $\Phi_{aI}$ for the gravitational sector [neglecting the last term of~\eqref{ktP}], the action \eqref{S_integrated} becomes
	\begin{eqnarray}
		\label{S_hf}
		S & = & \int_{\mathbb{R} \times \Sigma} dtd^{3}x \bigg[  2 \kappa \tilde{\Pi}^{aI} \dot{\Phi}_{aI} + \dfrac{1}{2} n_{I} \Big( \bar{\phi} \gamma^{I} E \dot{\phi} \notag \\
		& &  -  \dot{\bar{\phi}} \gamma^{I} E^{\dagger} \phi \Big) - \lambda_{IJ} \tilde{\mathcal{G}}^{IJ} - 2 N^{a} \tilde{\mathcal{D}}_{a} - \uac{N} \tilde{\tilde{\mathcal{H}}} \bigg],
	\end{eqnarray}
	where the first-class constraints are given by  
	\begin{widetext}
		\begin{subequations} 
			\label{const_nc_AV_hf}
			\begin{eqnarray}
				\label{const_nc_AV_hf_G}
				\tilde{\mathcal{G}}^{IJ} & = & 2 \kappa \Big( \tilde{\Pi}^{a[I} \Phi_{a}{}^{J]} + 2 P^{IJ}{}_{KL} \tilde{\Pi}^{a[K} n^{M]} \Gamma_{a}{}^{L}{}_{M} \Big) 
				+ \dfrac{1}{2} n^{[I} \left( \theta \tilde{V}^{J]} + \xi \tilde{A}^{J]} \right) + \dfrac{1}{4} \epsilon^{IJ}{}_{KL} n^{K} \tilde{A}^{L},  \\
				\label{const_nc_AV_hf_D}
				\tilde{\mathcal{D}}_{a} & = &  \kappa \Big( 2 \tilde{\Pi}^{bI} \partial_{[a} \Phi_{b]I} - \Phi_{aI} \partial_{b} \tilde{\Pi}^{bI} \Big) + \frac{1}{4} n_{I} \left( \bar{\phi} \gamma^{I} \partial_{a} \phi - \partial_{a} \bar{\phi} \gamma^{I} \phi \right) - \dfrac{1}{4} \Gamma_{aIJ} n^{I} \left( \theta \tilde{V}^{J} + \xi \tilde{A}^{J} \right) , \\
				\tilde{\tilde{\mathcal{H}}} & = & \kappa \tilde{\Pi}^{aI} \tilde{\Pi}^{bJ} R_{abIJ} + 2 \kappa \tilde{\Pi}^{a[I} \tilde{\Pi}^{|b|J]} \bigg[ \Phi_{aI} \Phi_{bJ} + 2 \Phi_{aI} \ovg{\Gamma}_{bJK} n^{K} + \frac{1}{\gamma^{2}} q^{KL} \Gamma_{aIK} \Gamma_{bJL} + \bigg( \Gamma_{aIK} + \dfrac{2}{\gamma} \ast \Gamma_{aIK} \bigg) \notag \\
				& &\times \Gamma_{bJL} n^{K} n^{L} \bigg] + \dfrac{1}{2}  \tilde{\Pi}^{aI} \left( \bar{\phi} \gamma_{I} \nabla_{a} \phi - \overline{\nabla_{a} \phi} \gamma_{I} \phi \right) + \dfrac{1}{2} \epsilon^{IJ}{}_{KL} \Big( \Phi_{aI} + \ovg{\Gamma}_{aIM} n^{M} \Big) n_{J}  \tilde{\Pi}^{aK} \tilde{A}^{L} \notag \\
				&& + \dfrac{3}{32 \kappa} \bigg\lbrace  q_{IJ} \tilde{A}^{I} \tilde{A}^{J} - \dfrac{\gamma^{2}}{\gamma^{2} + 1 } \bigg[- 2 \theta \left(  \xi + \frac{1}{\gamma} \right) \tilde{V}_{I} \tilde{A}^{I} - \theta^{2} \tilde{V}_{I} \tilde{V}^{I} + \left( 1 - \xi^{2} - 2 \frac{\xi}{\gamma}  \right) \tilde{A}_{I} \tilde{A}^{I} \bigg] \bigg\rbrace   \notag \\
				\label{const_nc_AV_hf_H}
				& &  - 2 h^{1/2}  \kappa \Lambda  -  h^{1/4}  m \bar{\phi} \phi.
			\end{eqnarray}
		\end{subequations}
	\end{widetext}
	\noindent
	and where we have also introduced the densitized fermion currents
	\begin{eqnarray}
		\label{Vd_def}
		\tilde{V}^{I} &:=& i \bar{\phi} \gamma^{I} \phi = h^{1/4} V^{I}, \\
		\label{Ad_def}
		\tilde{A}^{I} &:=& i \bar{\phi} \gamma^{5} \gamma^{I} \phi = h^{1/4} A^{I}, 
	\end{eqnarray}
	and we have written the covariant derivatives as
	\begin{subequations}
		\begin{eqnarray}
			\label{n_phi}
			\nabla_{a} \phi &:=& \partial_{a} \phi - \dfrac{1}{2} \Gamma^{b}{}_{ba} \phi + \dfrac{1}{2} \Gamma_{aIJ} \sigma^{IJ} \phi, \\
			\label{n_bphi}
			\overline{\nabla_{a} \phi} &:=& \partial_{a} \bar{\phi} - \dfrac{1}{2} \Gamma^{b}{}_{ba} \bar{\phi} - \dfrac{1}{2} \Gamma_{aIJ} \bar{\phi} \sigma^{IJ}.
		\end{eqnarray}
	\end{subequations}
	
	We close this section by remarking that the symplectic structure of the manifestly Lorentz-invariant Hamiltonian formulation~\eqref{S_hf} involves explicitly the parameters $\theta$ and $\xi$ in the fermionic part of it through $E$ and $E^{\dagger}$. We recall that fermion fields couple nonminimally to gravity~\eqref{S} when any of these parameters is nonvanishing. Therefore, one of the appealing features of this Hamiltonian formulation is that parameters that mediate the fermion-gravity interaction are present in the symplectic structure. Nevertheless, this is not a trivial fact as it might appear because it is possible to give an alternative Hamiltonian formulation that does {\it not} involve these parameters in the symplectic structure. This is done in the following section.   
	
	\section{Alternative Hamiltonian formulation involving half-densitized fermion fields}\label{Sec_hf2}
	
	We start again from our original manifestly Lorentz-invariant Hamiltonian formulation given by the action~\eqref{S_integrated} and the constraints~\eqref{const_nc_AV_G}-\eqref{const_nc_AV_H}. Like the Hamiltonian formulation reported in Sec.~\ref{Sec_hf}, here we also use half-densitized fermion fields defined by \eqref{hf1} and \eqref{hf2}. Nevertheless, instead of the gravitational configuration variable $C_{aI}$ present in~\eqref{S_integrated}, we use as gravitational configuration variable $\varphi_{aI}$ that is related to $C_{aI}$ by
	\begin{eqnarray}
		\varphi_{aI} := C_{aI} + \dfrac{1}{16 \kappa} \uac{\uac{h}}_{ab} \left( \tilde{\Pi}^{b}{}_{I} n_{J} - 2  \tilde{\Pi}^{b}{}_{J} n_{I} \right) \bar{\phi} \gamma^{J} \left(E - E^{\dagger}\right) \phi. \notag \\
	\end{eqnarray}
	Note that $\varphi_{aI}$ is also real. In terms of $\varphi_{aI}$, $\phi$, and ${\bar \phi}$, the symplectic structure of the action~\eqref{S_integrated} acquires the form
	\begin{eqnarray}
		\label{ase}
		&& 2 \kappa \tilde{\Pi}^{aI} \dot{C}_{aI} + \dfrac{1}{2} h^{1/4} n_{I} \Big( \bar{\psi} \gamma^{I} E \dot{\psi} -  \dot{\bar{\psi}} \gamma^{I} E^{\dagger} \psi \Big) \notag \\
		&=& 2 \kappa \tilde{\Pi}^{aI} \dot{\varphi}_{aI} + \dfrac{1}{2} n_{I} \Big(  \bar{\phi} \gamma^{I} \dot{\phi} - \dot{\bar{\phi}} \gamma^{I}  \phi \Big) \notag \\
		& & - \dfrac{1}{8} \partial_{t} \left[ n_{I} \bar{\phi} \gamma^{I} \left( E - E^{\dagger} \right)  \phi \right]. 
	\end{eqnarray}
	Some remarks are in order. First, due to the fact that the original symplectic structure~\eqref{S_integrated} is real and the last line of~\eqref{ase} is also real, the terms in the second line of~\eqref{ase} define a real symplectic structure if its last line is dropped. Second, note that this resulting symplectic structure does {\it not} involve the parameters $\theta$ and $\xi$. Therefore, it is remarkable that these parameters that mediate the nonminimal coupling of fermions fields to the gravitational field have been removed from the symplectic structure through a redefinition of the gravitational configuration variable and the use of half-densitized fermion fields.
	
	Thus, the action~\eqref{S_integrated} becomes
	\begin{eqnarray}
		\label{S_hf2}
		S & = & \int_{\mathbb{R} \times \Sigma} dtd^{3}x \bigg[  2 \kappa \tilde{\Pi}^{aI} \dot{\varphi}_{aI} + \dfrac{1}{2} n_{I} \Big( \bar{\phi} \gamma^{I} \dot{\phi}  -  \dot{\bar{\phi}} \gamma^{I}  \phi \Big) \notag \\
		& & - \lambda_{IJ} \tilde{\mathcal{G}}^{IJ} - 2 N^{a} \tilde{\mathcal{D}}_{a} - \uac{N} \tilde{\tilde{\mathcal{H}}} \bigg],
	\end{eqnarray}
	and the first-class constraints are given by
	\begin{widetext}
		\begin{subequations} 
			\begin{eqnarray}
				\tilde{\mathcal{G}}^{IJ} & = & 2 \kappa \Big( \tilde{\Pi}^{a[I} \varphi_{a}{}^{J]} + 2 P^{IJ}{}_{KL} \tilde{\Pi}^{a[K} n^{M]} \Gamma_{a}{}^{L}{}_{M} \Big)  + \dfrac{1}{4} \epsilon^{IJ}{}_{KL} n^{K} \tilde{A}^{L},  \\
				\tilde{\mathcal{D}}_{a} & = &  \kappa \Big( 2 \tilde{\Pi}^{bI} \partial_{[a} \varphi_{b]I} - \varphi_{aI} \partial_{b} \tilde{\Pi}^{bI} \Big) + \frac{1}{4} n_{I} \left( \bar{\phi} \gamma^{I} \partial_{a} \phi - \partial_{a} \bar{\phi} \gamma^{I} \phi \right)  , \\
				\tilde{\tilde{\mathcal{H}}} & = & \kappa \tilde{\Pi}^{aI} \tilde{\Pi}^{bJ} R_{abIJ} + 2 \kappa \tilde{\Pi}^{a[I} \tilde{\Pi}^{|b|J]} \bigg[ \varphi_{aI} \varphi_{bJ} + 2 \varphi_{aI} \ovg{\Gamma}_{bJK} n^{K} + \frac{1}{\gamma^{2}} q^{KL} \Gamma_{aIK} \Gamma_{bJL} + \bigg( \Gamma_{aIK} + \dfrac{2}{\gamma} \ast \Gamma_{aIK} \bigg) \notag \\
				& &\times \Gamma_{bJL} n^{K} n^{L} \bigg] + \dfrac{1}{2}  \tilde{\Pi}^{aI} \left( \bar{\phi} \gamma_{I} \nabla_{a} \phi - \overline{\nabla_{a} \phi} \gamma_{I} \phi \right) + \dfrac{1}{2} \epsilon^{IJ}{}_{KL} \Big( \varphi_{aI} + \ovg{\Gamma}_{aIM} n^{M} \Big) n_{J}  \tilde{\Pi}^{aK} \tilde{A}^{L} \notag \\
				&& + \dfrac{3}{32 \kappa} \bigg\lbrace  q_{IJ} \tilde{A}^{I} \tilde{A}^{J} - \dfrac{\gamma^{2}}{\gamma^{2} + 1 } \bigg[- 2 \theta \left(  \xi + \frac{1}{\gamma} \right) \tilde{V}_{I} \tilde{A}^{I} - \theta^{2} \tilde{V}_{I} \tilde{V}^{I} + \left( 1 - \xi^{2} - 2 \frac{\xi}{\gamma}  \right) \tilde{A}_{I} \tilde{A}^{I} \bigg] \bigg\rbrace   \notag \\
				& &  - 2 h^{1/2}  \kappa \Lambda  -  h^{1/4}  m \bar{\phi} \phi.
			\end{eqnarray}
		\end{subequations}
	\end{widetext}

	Note also that the coupling parameters $\theta$ and $\xi$ only appear in the Hamiltonian constraint $\tilde{\tilde{\mathcal{H}}}$. As the reader can notice, the manifestly Lorentz-invariant Hamiltonian formulation~\eqref{S_hf2} can also be obtained from the formulation~\eqref{S_hf}, contained in Sec.~\ref{Sec_hf}, and the fact that the variables $\Phi_{aI}$ of such a formulation and $\varphi_{aI}$ are related by  
	\begin{eqnarray}
		\label{rel_pp}
		\varphi_{aI} = \Phi_{aI} - \dfrac{1}{8 \kappa} \uac{\uac{h}}_{ab}  n_{I} \tilde{\Pi}^{b}{}_{J} \bar{\phi} \gamma^{J} \left(E - E^{\dagger}\right) \phi.
	\end{eqnarray}

	\section{Time gauge}\label{Sec_tg}
	The Hamiltonian formulations involving only first-class constraints reported in Secs.~\ref{Sec_Ham_form}--\ref{Sec_hf2} are manifestly Lorentz invariant, i.e., they are invariant under local $SO(3,1)$ gauge transformations, and thus, the local Lorentz symmetry has not been spoiled. Nevertheless, loop quantum gravity~\cite{RovBook,ThieBook,Ashtekar0407} employs Barbero's canonical variables, which are invariant under local $SO(3)$ gauge transformations. Therefore, motivated by loop quantum gravity, we now study the form of the Hamiltonian formulations of Secs.~\ref{Sec_Ham_form}--\ref{Sec_hf2} when the Lorentz group is broken down to its compact subgroup $SO(3)$ [and thus to its double cover $SU(2)$].
	
	The time gauge is given by
	\begin{subequations}
		\begin{eqnarray}
			\tilde{\Pi}^{a0} &=& 0, \label{GC} \\
			\tilde{\mathcal{G}}^{i0} &=& 0. \label{GC2}
		\end{eqnarray}
	\end{subequations}
	Note that $\tilde{\Pi}^{a0}=0$ is equivalent to $n_{i}=0$ ($i=1,2,3$) as long as $\det(\tilde{\Pi}^{ai}) \neq 0$, which is assumed throughout this section. From the explicit form of $n_{I}$ given in~\eqref{nI}, we observe that the only nonzero component of $n_{I}$ is $n_{0} = \mbox{sgn}[\det(\tilde{\Pi}^{ai})]$.
	
	We now fix some notation that will be used in what follows. From~\eqref{CD} and~\eqref{GC} we have $\Gamma_{a0i} = 0$ and define
	\begin{equation}
		\label{SC}
		\Gamma_{ai} := - \dfrac{1}{2} \epsilon_{ijk} \Gamma_{a}{}^{jk}, 
	\end{equation}
	where $\Gamma_{a}{}^i{}_j$ is the spin connection on $\Sigma$. The curvature of $\Gamma_{ai}$ is
	\begin{equation}
		\label{Rabi}
		R_{abi} := - \frac{1}{2} \epsilon_{ijk} R_{ab}{}^{jk} = \partial_{a} \Gamma_{bi} - \partial_{b} \Gamma_{ai} + \epsilon_{ijk} \Gamma_{a}{}^{j} \Gamma_{b}{}^{k},
	\end{equation} 
	and, from~\eqref{CD}, we recall that
	\begin{equation}
		\label{CDtg}
		\nabla_{a} \tilde{\Pi}^{bi} = \partial_{a} \tilde{\Pi}^{bi} + \Gamma^{b}{}_{ca} \tilde{\Pi}^{ci} - \Gamma^{c}{}_{ca} \tilde{\Pi}^{bi} + \epsilon^{i}{}_{jk} \Gamma_{a}{}^{j} \tilde{\Pi}^{bk} =0.
	\end{equation}
	
	\subsection{Time gauge in the Hamiltonian formulation of Sec.~\ref{Sec_Ham_form}}
	We impose the time gauge on the action~\eqref{S_integrated} with the constraints from~\eqref{const_nc_AV_G} to \eqref{const_nc_AV_H}. Solving $\tilde{\mathcal{G}}^{i0}=0$ using~\eqref{const_nc_AV_G}, we get
	\begin{equation}
		C_{a0} = n_{0} \uac{\Pi}_{ai} \partial_{b} \tilde{\Pi}^{bi} + \dfrac{n_{0}}{4 \kappa} h^{1/4} \uac{\Pi}_{ai} \left( \theta V^{i} + \xi A^{i} \right).
	\end{equation}
	Using this and~\eqref{GC}, the action~\eqref{S_integrated} takes the form
	\begin{eqnarray}
		\label{S_tg_nc}
		S & = & \int_{\mathbb{R} \times \Sigma} dtd^{3}x \Big[  2 \kappa \tilde{\Pi}^{ai} \dot{C}_{ai} + \dfrac{1}{2} h^{1/4} n_{0} \Big( \bar{\psi} \gamma^{0} E \dot{\psi} \notag \\
		& &  -  \dot{\bar{\psi}} \gamma^{0} E^{\dagger} \psi \Big) - 2 \lambda_{i} \tilde{\mathcal{G}}^{i} - 2 N^{a} \tilde{\mathcal{D}}_{a} - \uac{N} \tilde{\tilde{\mathcal{H}}} \Big],
	\end{eqnarray}
	with
	\begin{widetext}
		\begin{subequations}
			\label{const_tg_nc}
			\begin{eqnarray}
				\tilde{\mathcal{G}}^{i} & = & \dfrac{\kappa n_{0}}{\gamma} \left[ \partial_{a} \tilde{\Pi}^{ai} + \epsilon^{i}{}_{jk} \left(n_{0} \gamma C_{a}{}^{j} \right) \tilde{\Pi}^{ak} \right]  + \dfrac{n_{0}}{4} h^{1/4} A^{i}, \\
				\tilde{\mathcal{D}}_{a} & = & \kappa \left( 2 \tilde{\Pi}^{bi} \partial_{[a} C_{b]i} - C_{ai} \partial_{b} \tilde{\Pi}^{bi} \right) + \dfrac{n_{0}}{4} h^{1/4}  \Big[  \bar{\psi} \gamma^{0} \partial_{a} \psi - \partial_{a} \bar{\psi} \gamma^{0}  \psi   + \partial_{a} \left( \theta V^{0} + \xi A^{0} \right) \Big], \\
				\tilde{\tilde{\mathcal{H}}} & = &  - \kappa \epsilon_{ijk} \tilde{\Pi}^{ai} \tilde{\Pi}^{bj} R_{ab}{}^{k} + 2 \kappa \tilde{\Pi}^{a[i} \tilde{\Pi}^{|b|j]} \left( C_{ai} - \dfrac{n_{0}}{\gamma} \Gamma_{ai} \right) \left( C_{bj} - \dfrac{n_{0}}{\gamma} \Gamma_{bj} \right) + \dfrac{1}{2} h^{1/4} \tilde{\Pi}^{ai} \left( \bar{\psi} \gamma_{i} \nabla_{a} \psi - \overline{\nabla_{a} \psi} \gamma_{i} \psi \right) \notag \\
				& &  + \dfrac{n_{0}}{2} h^{1/4}  \left( C_{ai} - \dfrac{n_{0}}{\gamma} \Gamma_{ai} \right) \left[ \tilde{\Pi}^{ai} \left( \theta V^{0} + \xi A^{0} \right) + \epsilon^{i}{}_{jk} \tilde{\Pi}^{aj} A^{k} \right]   + \frac{3}{32 \kappa} h^{1/2} \bigg\lbrace  A_{i} A^{i} +  \left( \theta V^{0} +  \xi  A^{0} \right)^{2} \notag \\
				& & - \dfrac{\gamma^{2}}{\gamma^{2} + 1 } \bigg[ - 2 \theta \left(  \xi + \frac{1}{\gamma} \right) V_{I}A^{I}  - \theta^{2} V_{I} V^{I} + \left( 1 - \xi^{2} - 2 \frac{\xi}{\gamma}  \right) A_{I} A^{I} \bigg] \bigg\rbrace  - h^{1/2} \left( 2 \kappa \Lambda +  m  \bar{\psi} \psi  \right),
			\end{eqnarray}
		\end{subequations}
	\end{widetext}
	where we have defined $\lambda_{i}:= - (1/2) \epsilon_{ijk} \lambda^{jk}$, $\tilde{\mathcal{G}}^{i} := -(1/2) \epsilon^{i}{}_{jk} \tilde{\mathcal{G}}^{jk}$, and
	\begin{subequations}
		\begin{eqnarray}
			\nabla_{a} \psi & = & \partial_{a} \psi - \dfrac{1}{2} \epsilon_{ijk} \Gamma_{a}{}^{i} \sigma^{jk} \psi, \\
			\overline{\nabla_{a} \psi} & = & \partial_{a} \bar{\psi} + \dfrac{1}{2} \epsilon_{ijk} \Gamma_{a}{}^{i} \bar{\psi} \sigma^{jk}.
		\end{eqnarray}
	\end{subequations}
	[See \eqref{n_psi} and \eqref{n_bpsi}]. From $\tilde{\mathcal{G}}^{i}$, we identify the $SO(3)$ connection,
	\begin{equation}
		\label{A_def}
		A_{ai} := n_{0} \gamma C_{ai},
	\end{equation}
	and we define its field strength as
	\begin{equation}
		\label{Fabi}
		F_{abi} := \partial_{a} A_{bi}  - \partial_{b} A_{ai} + \epsilon_{ijk} A_{a}{}^{j} A_{b}{}^{k}.
	\end{equation}
	[See~\eqref{relation_connections} below to conclude alternatively that $A_{ai}$ is indeed a connection.] Using the identity (41) of Ref.~\cite{Montesinos2004a} and integrating by parts, which requires to replace $\lambda_i$ with $\mu_i$ given by $\mu_{i}:= \lambda_{i} - (n_{0}/\gamma) \tilde{\Pi}^{a}{}_{i} \nabla_{a} \uac{N}$, we obtain the action
	\begin{eqnarray}
		\label{S_tg_nc_A2}
		S & = & \int_{\mathbb{R} \times \Sigma} dtd^{3}x \bigg[ 2 \dfrac{\kappa n_{0}}{\gamma} \tilde{\Pi}^{ai} \dot{A}_{ai} + \dfrac{1}{2} h^{1/4} n_{0} \Big( \bar{\psi} \gamma^{0} E \dot{\psi} \notag \\
		& &  -  \dot{\bar{\psi}} \gamma^{0} E^{\dagger} \psi \Big) - 2 \mu_{i} \tilde{\mathcal{G}}^{i} - 2 N^{a} \tilde{\mathcal{D}}_{a} - \uac{N} \tilde{\tilde{\mathcal{C}}} \bigg],
	\end{eqnarray}
	where the first-class constraints are given by
	\begin{widetext}
		\begin{subequations}
			\label{const_tg_nc_A2}
			\begin{eqnarray}
				\label{const_tg_nc_A2_G}
				\tilde{\mathcal{G}}^{i} & = & \dfrac{\kappa n_{0}}{\gamma} \left[ \partial_{a} \tilde{\Pi}^{ai} + \epsilon^{i}{}_{jk}  A_{a}{}^{j} \tilde{\Pi}^{ak} \right]  + \dfrac{n_{0}}{4}  h^{1/4} A^{i}, \\
				\label{const_tg_nc_A2_Ca}
				\tilde{\mathcal{D}}_{a} & = & \dfrac{\kappa n_{0}}{\gamma} \left( 2 \tilde{\Pi}^{bi} \partial_{[a} A_{b]i} - A_{ai} \partial_{b} \tilde{\Pi}^{bi} \right) + \dfrac{n_{0}}{4} h^{1/4}  \Big[  \bar{\psi} \gamma^{0} \partial_{a} \psi - \partial_{a} \bar{\psi} \gamma^{0}  \psi   + \partial_{a} \left( \theta V^{0} + \xi A^{0} \right) \Big],\\
				\tilde{\tilde{\mathcal{C}}} & = &  \dfrac{\kappa}{\gamma^{2}} \epsilon_{ijk} \tilde{\Pi}^{ai} \tilde{\Pi}^{bj} \left[F_{ab}{}^{k} - \left( 1 + \gamma^{2} \right) R_{ab}{}^{k} \right] + \dfrac{1}{2} h^{1/4} \tilde{\Pi}^{ai} \left[ \bar{\psi} \left( \mathds{1} - \dfrac{\mathrm{i}}{\gamma} \gamma^{5} \right) \gamma_{i} \nabla_{a} \psi - \overline{\nabla_{a} \psi} \left( \mathds{1} + \dfrac{\mathrm{i}}{\gamma} \gamma^{5} \right) \gamma_{i} \psi \right] \notag \\
				& &  + \dfrac{1}{2\gamma} h^{1/4}  \left( A_{ai} - \Gamma_{ai} \right) \left[ \tilde{\Pi}^{ai} \left( \theta V^{0} + \xi A^{0} \right) +  \epsilon^{i}{}_{jk} \tilde{\Pi}^{aj} A^{k} \right]  + \frac{3}{32 \kappa} h^{1/2} \bigg\lbrace   A_{i} A^{i} +  \left( \theta V^{0} +  \xi  A^{0} \right)^{2} \notag \\
				\label{const_tg_nc_A2_C}
				& & - \dfrac{\gamma^{2}}{\gamma^{2} + 1 }  \bigg[ - 2 \theta \left(  \xi + \frac{1}{\gamma} \right) V_{I}A^{I}  - \theta^{2} V_{I} V^{I} + \left( 1 - \xi^{2} - 2 \frac{\xi}{\gamma}  \right) A_{I} A^{I} \bigg] \bigg\rbrace - h^{1/2} \left( 2 \kappa \Lambda +  m \bar{\psi} \psi  \right). 
			\end{eqnarray}
		\end{subequations}
	\end{widetext}

	This is the desired Hamiltonian formulation coming from~\eqref{S_integrated} when the time gauge is imposed on.
	
	
	\subsection{Time gauge in the Hamiltonian formulation of Sec.~\ref{Sec_hf}}
	
	We impose the time gauge on the Hamiltonian formulation given in the action~\eqref{S_hf} with the constraints from~\eqref{const_nc_AV_hf_G} to \eqref{const_nc_AV_hf_H}. Solving $\tilde{\mathcal{G}}^{i0}=0$ using~\eqref{const_nc_AV_hf_G}, we get
	\begin{equation}
		\Phi_{a0} = n_{0} \uac{\Pi}_{ai} \partial_{b} \tilde{\Pi}^{bi} + \dfrac{n_{0}}{4 \kappa} \uac{\Pi}_{ai} \left( \theta \tilde{V}^{i} + \xi \tilde{A}^{i} \right).
	\end{equation}
	Using this and~\eqref{GC}, the action~\eqref{S_hf} takes the form
	\begin{eqnarray}
		\label{S_tg_nc_hf}
		S & = & \int_{\mathbb{R} \times \Sigma} dtd^{3}x \Big[  2 \kappa \tilde{\Pi}^{ai} \dot{\Phi}_{ai} + \dfrac{1}{2} n_{0} \Big( \bar{\phi} \gamma^{0} E \dot{\phi} \notag \\
		& &  -  \dot{\bar{\phi}} \gamma^{0} E^{\dagger} \phi \Big) - 2 \lambda_{i} \tilde{\mathcal{G}}^{i} - 2 N^{a} \tilde{\mathcal{D}}_{a} - \uac{N} \tilde{\tilde{\mathcal{H}}} \Big],
	\end{eqnarray}
	with
	\begin{widetext}
		\begin{subequations}
			\label{const_tg_nc_hf}
			\begin{eqnarray}
				\tilde{\mathcal{G}}^{i} & = & \dfrac{\kappa n_{0}}{\gamma} \left[ \partial_{a} \tilde{\Pi}^{ai} + \epsilon^{i}{}_{jk} \left(n_{0} \gamma \Phi_{a}{}^{j} \right) \tilde{\Pi}^{ak} \right]  + \dfrac{n_{0}}{4} \tilde{A}^{i}, \\
				\tilde{\mathcal{D}}_{a} & = & \kappa \left( 2 \tilde{\Pi}^{bi} \partial_{[a} \Phi_{b]i} - \Phi_{ai} \partial_{b} \tilde{\Pi}^{bi} \right) + \dfrac{n_{0}}{4} \big( \bar{\phi} \gamma^{0} \partial_{a} \phi - \partial_{a} \bar{\phi} \gamma^{0}  \phi \big), \\
				\tilde{\tilde{\mathcal{H}}} & = &  - \kappa \epsilon_{ijk} \tilde{\Pi}^{ai} \tilde{\Pi}^{bj} R_{ab}{}^{k} + 2 \kappa \tilde{\Pi}^{a[i} \tilde{\Pi}^{|b|j]} \left( \Phi_{ai} - \dfrac{n_{0}}{\gamma} \Gamma_{ai} \right) \left( \Phi_{bj} - \dfrac{n_{0}}{\gamma} \Gamma_{bj} \right) + \dfrac{1}{2}  \tilde{\Pi}^{ai} \left( \bar{\phi} \gamma_{i} \nabla_{a} \phi - \overline{\nabla_{a} \phi} \gamma_{i} \phi \right) \notag \\
				& &  + \dfrac{n_{0}}{2} \epsilon_{ijk} \left( \Phi_{a}{}^{i} - \dfrac{n_{0}}{\gamma} \Gamma_{a}{}^{i} \right)  \tilde{\Pi}^{aj} \tilde{A}^{k}   + \frac{3}{32 \kappa} \bigg\lbrace   \tilde{A}_{i} \tilde{A}^{i}  - \dfrac{\gamma^{2}}{\gamma^{2} + 1 } \bigg[ - 2 \theta \left(  \xi + \frac{1}{\gamma} \right) \tilde{V}_{I} \tilde{A}^{I}  - \theta^{2} \tilde{V}_{I} \tilde{V}^{I} \notag \\
				& & + \left( 1 - \xi^{2} - 2 \frac{\xi}{\gamma}  \right) \tilde{A}_{I} \tilde{A}^{I} \bigg] \bigg\rbrace  - 2 h^{1/2} \kappa \Lambda -  m h^{1/4} \bar{\phi} \phi,
			\end{eqnarray}
		\end{subequations}
	\end{widetext}
	where we defined $\lambda_{i}:= - (1/2) \epsilon_{ijk} \lambda^{jk}$, $\tilde{\mathcal{G}}^{i} := -(1/2) \epsilon^{i}{}_{jk} \tilde{\mathcal{G}}^{jk}$, and 
	\begin{subequations}
		\begin{eqnarray}
			\nabla_{a} \phi &=& \partial_{a} \phi - \dfrac{1}{2} \Gamma^{b}{}_{ba} \phi - \dfrac{1}{2} \epsilon_{ijk} \Gamma_{a}{}^{i} \sigma^{jk} \phi, \\
			\overline{\nabla_{a} \phi} &=& \partial_{a} \bar{\phi} - \dfrac{1}{2} \Gamma^{b}{}_{ba} \bar{\phi}  + \dfrac{1}{2} \epsilon_{ijk} \Gamma_{a}{}^{i}  \bar{\phi} \sigma^{jk}.
		\end{eqnarray}
	\end{subequations}
	[See \eqref{n_phi} and \eqref{n_bphi}]. From $\tilde{\mathcal{G}}^{i}$, we identify immediately the connection
	\begin{equation}
		\label{A2_def} 
		\mathcal{A}_{ai} := n_{0} \gamma \Phi_{ai},
	\end{equation}
	and we define its field strength as
	\begin{equation}
		\label{Fabi2}
		\mathcal{F}_{abi} := \partial_{a} \mathcal{A}_{bi}  - \partial_{b} \mathcal{A}_{ai} + \epsilon_{ijk} \mathcal{A}_{a}{}^{j} \mathcal{A}_{b}{}^{k}.
	\end{equation}
	After an integration by parts that requires to replace $\lambda_i$ with $\mu_i$ given by $\mu_{i}:= \lambda_{i} - (n_{0}/\gamma) \tilde{\Pi}^{a}{}_{i} \nabla_{a} \uac{N}$, we obtain the action
	\begin{eqnarray}
		\label{S_tg_nc_hf_A2}
		S & = & \int_{\mathbb{R} \times \Sigma} dtd^{3}x \bigg[ 2 \dfrac{\kappa n_{0}}{\gamma} \tilde{\Pi}^{ai} \dot{\mathcal{A}}_{ai} + \dfrac{1}{2}  n_{0} \Big( \bar{\phi} \gamma^{0} E \dot{\phi} \notag \\
		& &  -  \dot{\bar{\phi}} \gamma^{0} E^{\dagger} \phi \Big) - 2 \mu_{i} \tilde{\mathcal{G}}^{i} - 2 N^{a} \tilde{\mathcal{D}}_{a} - \uac{N} \tilde{\tilde{\mathcal{C}}} \bigg],
	\end{eqnarray}
	where the first-class constraints are given by
	\begin{subequations}
		\label{const_tg_nc_hf_A2}
		\begin{eqnarray}
			\label{const_tg_nc_hf_A2_G}
			\tilde{\mathcal{G}}^{i} & = & \dfrac{\kappa n_{0}}{\gamma} \left[ \partial_{a} \tilde{\Pi}^{ai} + \epsilon^{i}{}_{jk}  \mathcal{A}_{a}{}^{j} \tilde{\Pi}^{ak} \right]  + \dfrac{n_{0}}{4}  \tilde{A}^{i}, \\
			\label{const_tg_nc_hf_A2_Ca}
			\tilde{\mathcal{D}}_{a} & = & \dfrac{\kappa n_{0}}{\gamma} \left( 2 \tilde{\Pi}^{bi} \partial_{[a} \mathcal{A}_{b]i} - \mathcal{A}_{ai} \partial_{b} \tilde{\Pi}^{bi} \right) \notag \\
				&& + \dfrac{n_{0}}{4} \big( \bar{\phi} \gamma^{0} \partial_{a} \phi - \partial_{a} \bar{\phi} \gamma^{0}  \phi \big),
		\end{eqnarray}
		\begin{widetext}
			\begin{eqnarray}	
				\tilde{\tilde{\mathcal{C}}} & = &  \dfrac{\kappa}{\gamma^{2}} \epsilon_{ijk} \tilde{\Pi}^{ai} \tilde{\Pi}^{bj} \left[ \mathcal{F}_{ab}{}^{k} - \left( 1 + \gamma^{2} \right) R_{ab}{}^{k} \right] + \dfrac{1}{2} \tilde{\Pi}^{ai} \left[ \bar{\phi} \left( \mathds{1} - \dfrac{\mathrm{i}}{\gamma} \gamma^{5} \right) \gamma_{i} \nabla_{a} \phi - \overline{\nabla_{a} \phi} \left( \mathds{1} + \dfrac{\mathrm{i}}{\gamma} \gamma^{5} \right) \gamma_{i} \phi \right] \notag \\
				& &  + \dfrac{1}{2\gamma} \epsilon_{ijk} \left( \mathcal{A}_{a}{}^{i} - \Gamma_{a}{}^{i} \right) \tilde{\Pi}^{aj} \tilde{A}^{k} + \frac{3}{32 \kappa} \bigg\lbrace   \tilde{A}_{i} \tilde{A}^{i}   - \dfrac{\gamma^{2}}{\gamma^{2} + 1 }  \bigg[ - 2 \theta \left(  \xi + \frac{1}{\gamma} \right) \tilde{V}_{I} \tilde{A}^{I}  - \theta^{2} \tilde{V}_{I} \tilde{V}^{I} \notag \\
				\label{const_tg_nc_hf_A2_C}
				& & + \left( 1 - \xi^{2} - 2 \frac{\xi}{\gamma} \right) \tilde{A}_{I} \tilde{A}^{I} \bigg] \bigg\rbrace - 2 h^{1/2} \kappa \Lambda -  m  h^{1/4} \bar{\phi} \phi. 
			\end{eqnarray}
		\end{widetext}	
	\end{subequations}

	Some remarks follow: first, note that in the Hamiltonian formulation given in the action~\eqref{S_tg_nc_hf_A2}, the coupling parameters $\theta$ and $\xi$ appear explicitly in the scalar constraint, specifically in the terms involving the fermion interaction of order fourth [cf.~\eqref{S_int}]. Second, the connection $\mathcal{A}_{ai}$ is related generically to the connection $A_{ai}$~\eqref{A_def} of the Hamiltonian formulation of Sec.~\ref{Sec_Ham_form} through the expression
	\begin{equation}
		\label{relation_connections}
		\mathcal{A}_{ai} = A_{ai} + \dfrac{\gamma}{8 \kappa} \uac{\Pi}_{ai} \left( \theta \tilde{V}^{0} + \xi \tilde{A}^{0} \right).
	\end{equation}
	Hence, in the minimal coupling of fermions ($\theta = \xi =0$), we have $\mathcal{A}_{ai} = A_{ai}$. In the nonminimal coupling of fermions (either $\theta \neq 0$ or $\xi \neq 0$, or both are different from zero), we have $\mathcal{A}_{ai} \neq A_{ai}$.    
	
	\subsection{Time gauge in the Hamiltonian formulation of Sec.~\ref{Sec_hf2}}
	
	Following the same approach as in the previous subsections, we impose the time gauge on the formulation given by the action~\eqref{S_hf2}. From $\mathcal{G}^{i0}=0$ we have
	\begin{equation}
		\varphi_{a0} = n_{0} \uac{\Pi}_{ai} \partial_{b} \tilde{\Pi}^{bi}.
	\end{equation}
	Then, identifying 
	\begin{equation}
		\label{A2_def2} 
		\mathcal{A}_{ai} := n_{0} \gamma \varphi_{ai},
	\end{equation}
	we arrive at the action
	\begin{eqnarray}
		\label{S_tg_nc_hf2_A2}
		S & = & \int_{\mathbb{R} \times \Sigma} dtd^{3}x \bigg[ 2 \dfrac{\kappa n_{0}}{\gamma} \tilde{\Pi}^{ai} \dot{\mathcal{A}}_{ai} + \dfrac{1}{2}  n_{0} \Big( \bar{\phi} \gamma^{0} \dot{\phi} -  \dot{\bar{\phi}} \gamma^{0} \phi \Big) \notag \\
		& &   - 2 \mu_{i} \tilde{\mathcal{G}}^{i} - 2 N^{a} \tilde{\mathcal{D}}_{a} - \uac{N} \tilde{\tilde{\mathcal{C}}} \bigg],
	\end{eqnarray}
	where $\mu_{i}:= - (1/2) \epsilon_{ijk} \lambda^{jk} - (n_{0}/\gamma) \tilde{\Pi}^{a}{}_{i} \nabla_{a} \uac{N}$, and  the constraints $\tilde{\mathcal{G}}^{i}$, $\tilde{\mathcal{D}}_{a}$, and $\tilde{\tilde{\mathcal{C}}}$ are the same as the expressions given in~\eqref{const_tg_nc_hf_A2_G}-\eqref{const_tg_nc_hf_A2_C}, respectively. Notice that the configuration variable $\mathcal{A}_{ai}$ is the same as the one defined in~\eqref{A2_def} since, in the time gauge, $\Phi_{ai} = \varphi_{ai}$ [see~\eqref{rel_pp}].

	\section{Symplectomorphism}\label{Sec_CT}
	So far, we have reported three manifestly Lorentz-invariant Hamiltonian formulations for the coupling of fermion fields to the Holst action described by the action principle~\eqref{S}, which are contained in Secs.~\ref{Sec_Ham_form}--\ref{Sec_hf2}; respectively. We have also reported the resulting Hamiltonian formulations once the time gauge is imposed on any of them, which are contained in Sec.~\ref{Sec_tg}. 
	
	Now, we report another manifestly Lorentz-invariant Hamiltonian formulation that is obtained from the formulation given by the action~\eqref{S_hf2} of Sec.~\ref{Sec_hf2} by means of a symplectomorphism that changes the gravitational configuration variable as
	\begin{eqnarray}
		\Psi_{aI} := \varphi_{aI} - W_{a}{}^{b}{}_{IJK} \ovg{\Gamma}_{b}\!{}^{JK},
	\end{eqnarray}
	whereas the remaining phase-space variables $\tilde{\Pi}^{aI}$, $\phi$, and ${\bar \phi}$ are left unchanged. That this transformation is indeed a symplectomorphism can be seen from the fact that
	\begin{eqnarray}
		&& 2 \kappa \tilde{\Pi}^{aI} \dot{\varphi}_{aI} + \dfrac{1}{2} n_{I} \Big( \bar{\phi} \gamma^{I} \dot{\phi}  -  \dot{\bar{\phi}} \gamma^{I}  \phi \Big) \notag \\
		& = &  2 \kappa \tilde{\Pi}^{aI} \dot{\Psi}_{aI} + \dfrac{1}{2} n_{I} \Big( \bar{\phi} \gamma^{I} \dot{\phi}  -  \dot{\bar{\phi}} \gamma^{I}  \phi \Big) + \kappa \partial_{a} \Big[ - 2  n_{I} \dot{\tilde{\Pi}}^{aI} \notag \\
		& &  - \dfrac{1}{\gamma} h^{1/2} \tilde{\eta}^{abc} \uac{\uac{h}}_{bd} \uac{\uac{h}}_{cf} \tilde{\Pi}^{f}{}_{I} \dot{\tilde{\Pi}}^{dI} \Big].
	\end{eqnarray}
	Thus, due to the fact that $\Sigma$ has no boundary, the new Hamiltonian formulation obtained from~\eqref{S_hf2} is given by the action
	\begin{eqnarray}
		\label{S_CT}
		S & = & \int_{\mathbb{R} \times \Sigma} dtd^{3}x \bigg[  2 \kappa \tilde{\Pi}^{aI} \dot{\Psi}_{aI} + \dfrac{1}{2} n_{I} \Big( \bar{\phi} \gamma^{I} \dot{\phi}  -  \dot{\bar{\phi}} \gamma^{I}  \phi \Big) \notag \\
		& & - \lambda_{IJ} \tilde{\mathcal{G}}^{IJ} - 2 N^{a} \tilde{\mathcal{D}}_{a} - \uac{N} \tilde{\tilde{\mathcal{H}}} \bigg],
	\end{eqnarray}
	where the first-class constraints are given by
	\begin{widetext}
		\begin{subequations} 
			\begin{eqnarray}
				\label{const_Psi_G}
				\tilde{\mathcal{G}}^{IJ} & = & 2 \kappa  \tilde{\Pi}^{a[I} \Psi_{a}{}^{J]}  + \dfrac{1}{4} \epsilon^{IJ}{}_{KL} n^{K} \tilde{A}^{L},  \\
				\label{const_Psi_D}
				\tilde{\mathcal{D}}_{a} & = &  \kappa \Big( 2 \tilde{\Pi}^{bI} \partial_{[a} \Psi_{b]I} - \Psi_{aI} \partial_{b} \tilde{\Pi}^{bI} \Big) + \frac{1}{4} n_{I} \left( \bar{\phi} \gamma^{I} \partial_{a} \phi - \partial_{a} \bar{\phi} \gamma^{I} \phi \right)  , \\
				\tilde{\tilde{\mathcal{H}}} & = & \kappa \tilde{\Pi}^{aI} \tilde{\Pi}^{bJ} R_{abIJ} + 2 \kappa \tilde{\Pi}^{a[I} \tilde{\Pi}^{|b|J]} \Psi_{aI} \Psi_{bJ} + \dfrac{1}{2}  \tilde{\Pi}^{aI} \left( \bar{\phi} \gamma_{I} \nabla_{a} \phi - \overline{\nabla_{a} \phi} \gamma_{I} \phi \right) + \dfrac{1}{2} \epsilon^{IJ}{}_{KL} \Psi_{aI} n_{J}  \tilde{\Pi}^{aK} \tilde{A}^{L} \notag \\
				&& + \dfrac{3}{32 \kappa} \bigg\lbrace q_{IJ} \tilde{A}^{I} \tilde{A}^{J} - \dfrac{\gamma^{2}}{\gamma^{2} + 1 } \bigg[- 2 \theta \left(  \xi + \frac{1}{\gamma} \right) \tilde{V}_{I} \tilde{A}^{I} - \theta^{2} \tilde{V}_{I} \tilde{V}^{I} + \left( 1 - \xi^{2} - 2 \frac{\xi}{\gamma}  \right) \tilde{A}_{I} \tilde{A}^{I} \bigg] \bigg\rbrace   \notag \\
				\label{const_Psi_H}
				& &  - 2 h^{1/2}  \kappa \Lambda  -  h^{1/4} m \bar{\phi} \phi.
			\end{eqnarray}
		\end{subequations}
	\end{widetext}

	Even though the coupling parameters $\theta$ and $\xi$ do not appear in the symplectic structure~\eqref{S_CT}, they appear in the Hamiltonian  constraint~\eqref{const_Psi_H}. Nevertheless, these constraints get a much simpler form by making two particular choices of the coupling parameters $\theta$ and $\xi$. Let us analyze them in what follows.
	
	\subsection{Fermion fields minimally coupled to the Palatini action}\label{choice1}
	When the coupling parameters are chosen as
	\begin{eqnarray}
		\xi = -\frac{1}{\gamma}, \quad \theta = 0,
	\end{eqnarray}
	which amounts to choose $E =  \mathds{1} + (\mathrm{i} /\gamma) \gamma^{5}$, the constraints~\eqref{const_Psi_G}-\eqref{const_Psi_H} become
	\begin{subequations} 
		\begin{eqnarray}
			\label{const_Psi_G_1}
			\tilde{\mathcal{G}}^{IJ} & = & 2 \kappa  \tilde{\Pi}^{a[I} \Psi_{a}{}^{J]}  + \dfrac{1}{4} \epsilon^{IJ}{}_{KL} n^{K} \tilde{A}^{L},  \\
			\label{const_Psi_D_1}
			\tilde{\mathcal{D}}_{a} & = &  \kappa \Big( 2 \tilde{\Pi}^{bI} \partial_{[a} \Psi_{b]I} - \Psi_{aI} \partial_{b} \tilde{\Pi}^{bI} \Big) \notag \\
			&&+ \frac{1}{4} n_{I} \left( \bar{\phi} \gamma^{I} \partial_{a} \phi - \partial_{a} \bar{\phi} \gamma^{I} \phi \right)  , \\
			\tilde{\tilde{\mathcal{H}}} & = & \kappa \tilde{\Pi}^{aI} \tilde{\Pi}^{bJ} R_{abIJ} + 2 \kappa \tilde{\Pi}^{a[I} \tilde{\Pi}^{|b|J]} \Psi_{aI} \Psi_{bJ}  \notag \\
			&&+ \dfrac{1}{2}  \tilde{\Pi}^{aI} \left( \bar{\phi} \gamma_{I} \nabla_{a} \phi - \overline{\nabla_{a} \phi} \gamma_{I} \phi \right)  \notag \\
			&& + \dfrac{1}{2} \epsilon^{IJ}{}_{KL} \Psi_{aI} n_{J}  \tilde{\Pi}^{aK} \tilde{A}^{L} + \dfrac{3}{32 \kappa} n_{I} n_{J} \tilde{A}^{I} \tilde{A}^{J} \notag \\
			\label{const_Psi_H_1}
			& &  - 2 h^{1/2}  \kappa \Lambda  - h^{1/4} m \bar{\phi} \phi.
		\end{eqnarray}
	\end{subequations}

	Notice that there is no Immirzi parameter in the first-class constraints at all. Therefore, the absence of the Immirzi parameter and the form of the constraints indicate that this manifestly Lorentz-invariant Hamiltonian formulation corresponds to fermion fields minimally coupled to the Palatini action~\cite{MRC_PalFer} (Einstein-Cartan-Dirac), i.e., this Hamiltonian formulation comes from the action given by~\eqref{S_F} (with $\theta=0$ and $\xi=0$) added to the Palatini action.

	\subsection{Fermion fields nonminimally coupled to the Palatini action}\label{choice2}
	When the coupling parameters are chosen as
	\begin{eqnarray}
		\xi = -\frac{1}{\gamma}, \quad \theta = \vartheta \sqrt{1 + \gamma^{-2}}, \quad \vartheta \in \mathbb{R},
	\end{eqnarray}
	which amounts to choose $E =  ( 1 + \mathrm{i} \vartheta \sqrt{1 + \gamma^{-2}}) \mathds{1} + (\mathrm{i} /\gamma ) \gamma^{5}$, the constraints~\eqref{const_Psi_G}-\eqref{const_Psi_H} become
	\begin{subequations} 
		\begin{eqnarray}
			\label{const_Psi_G_2}
			\tilde{\mathcal{G}}^{IJ} & = & 2 \kappa  \tilde{\Pi}^{a[I} \Psi_{a}{}^{J]}  + \dfrac{1}{4} \epsilon^{IJ}{}_{KL} n^{K} \tilde{A}^{L},  \\
			\label{const_Psi_D_2}
			\tilde{\mathcal{D}}_{a} & = &  \kappa \Big( 2 \tilde{\Pi}^{bI} \partial_{[a} \Psi_{b]I} - \Psi_{aI} \partial_{b} \tilde{\Pi}^{bI} \Big) \notag \\
			&&+ \frac{1}{4} n_{I} \left( \bar{\phi} \gamma^{I} \partial_{a} \phi - \partial_{a} \bar{\phi} \gamma^{I} \phi \right)  , \\
			\tilde{\tilde{\mathcal{H}}} & = & \kappa \tilde{\Pi}^{aI} \tilde{\Pi}^{bJ} R_{abIJ} + 2 \kappa \tilde{\Pi}^{a[I} \tilde{\Pi}^{|b|J]} \Psi_{aI} \Psi_{bJ}  \notag \\
			&&+ \dfrac{1}{2}  \tilde{\Pi}^{aI} \left( \bar{\phi} \gamma_{I} \nabla_{a} \phi - \overline{\nabla_{a} \phi} \gamma_{I} \phi \right)  \notag \\
			&& + \dfrac{1}{2} \epsilon^{IJ}{}_{KL} \Psi_{aI} n_{J}  \tilde{\Pi}^{aK} \tilde{A}^{L} + \dfrac{3}{32 \kappa} \Big( \vartheta^{2} \tilde{V}_{I} \tilde{V}^{I}  \notag \\
			\label{const_Psi_H_2}
			& & + n_{I} n_{J} \tilde{A}^{I} \tilde{A}^{J}\Big) - 2 h^{1/2}  \kappa \Lambda  -  h^{1/4} m \bar{\phi} \phi.
		\end{eqnarray}
	\end{subequations}

	This manifestly Lorentz-invariant Hamiltonian formulation involves $\vartheta$ as its single real coupling parameter, which is present in the vector-vector interaction of the Hamiltonian constraint~\eqref{const_Psi_H_2}. Note also that this formulation is invariant under parity transformations since the Hamiltonian constraint~\eqref{const_Psi_H_2} does not involve the vector-axial interaction. The only difference of this formulation with respect to the formulation given in Sec.~\ref{choice1} is precisely the term involving $\vartheta$. Like that, this formulation also does not involve the Immirzi parameter. Therefore, the absence of the Immirzi parameter, the form of the constraints, and the presence of the parameter $\vartheta$ indicate that this manifestly Lorentz-invariant Hamiltonian formulation corresponds to fermion fields nonminimally coupled to the Palatini action. More precisely, the Hamiltonian formulation given by the constraints~\eqref{const_Psi_G_2}-\eqref{const_Psi_H_2} comes from the action given by~\eqref{S_F} (with $\theta=\vartheta$ and $\xi=0$) added to the Palatini action~\cite{MRC_PalFer}.
	
	\subsection{Time gauge}\label{tg_symp}
	For the sake of completeness, we also impose the time gauge on the Hamiltonian formulation given by~\eqref{S_CT}. From $\tilde{\mathcal{G}}^{i0}=0$, we obtain
	\begin{equation}
		\Psi_{a0} = 0.
	\end{equation}
	Thus, under the time gauge, the action~\eqref{S_CT} becomes
	\begin{eqnarray}
		\label{S_CT_tg}
		S & = & \int_{\mathbb{R} \times \Sigma} dtd^{3}x \bigg[  2 \kappa \tilde{\Pi}^{ai} \dot{\Psi}_{ai} + \dfrac{1}{2} n_{0} \Big( \bar{\phi} \gamma^{0} \dot{\phi}  -  \dot{\bar{\phi}} \gamma^{0}  \phi \Big) \notag \\
		& & - 2 \lambda_{i} \tilde{\mathcal{G}}^{i} - 2 N^{a} \tilde{\mathcal{D}}_{a} - \uac{N} \tilde{\tilde{\mathcal{H}}} \bigg],
	\end{eqnarray}
	where we defined $\lambda_{i}:= - (1/2) \epsilon_{ijk} \lambda^{jk}$ and $\tilde{\mathcal{G}}^{i} := -(1/2) \epsilon^{i}{}_{jk} \tilde{\mathcal{G}}^{jk}$, and the first-class constraints are given by
	\begin{widetext}
		\begin{subequations} 
			\begin{eqnarray}
				\label{const_Psi_G_tg}
				\tilde{\mathcal{G}}^{i} & = &  \kappa \epsilon^{i}{}_{jk} \Psi_{a}{}^{j} \tilde{\Pi}^{ak}   + \dfrac{n_{0}}{4}   \tilde{A}^{i},  \\
				\label{const_Psi_D_tg}
				\tilde{\mathcal{D}}_{a} & = &  \kappa \Big( 2 \tilde{\Pi}^{bi} \partial_{[a} \Psi_{b]i} - \Psi_{ai} \partial_{b} \tilde{\Pi}^{bi} \Big) + \frac{n_{0}}{4}  \left( \bar{\phi} \gamma^{0} \partial_{a} \phi - \partial_{a} \bar{\phi} \gamma^{0} \phi \right)  , \\
				\tilde{\tilde{\mathcal{H}}} & = & - \kappa \epsilon_{ijk} \tilde{\Pi}^{ai} \tilde{\Pi}^{bj} R_{ab}{}^{k} + 2 \kappa \tilde{\Pi}^{a[i} \tilde{\Pi}^{|b|j]} \Psi_{ai} \Psi_{bj} + \dfrac{1}{2}  \tilde{\Pi}^{ai} \left( \bar{\phi} \gamma_{i} \nabla_{a} \phi - \overline{\nabla_{a} \phi} \gamma_{i} \phi \right) + \dfrac{n_{0}}{2}  \epsilon_{ijk} \Psi_{a}{}^{i}  \tilde{\Pi}^{aj} \tilde{A}^{k} \notag \\
				&& + \dfrac{3}{32 \kappa} \bigg\lbrace   \tilde{A}_{i} \tilde{A}^{i} - \dfrac{\gamma^{2}}{\gamma^{2} + 1 } \bigg[- 2 \theta \left(  \xi + \frac{1}{\gamma} \right) \tilde{V}_{I} \tilde{A}^{I} - \theta^{2} \tilde{V}_{I} \tilde{V}^{I} + \left( 1 - \xi^{2} - 2 \frac{\xi}{\gamma}  \right) \tilde{A}_{I} \tilde{A}^{I} \bigg] \bigg\rbrace   \notag \\
				\label{const_Psi_H_tg}
				& &  - 2 h^{1/2}  \kappa \Lambda  -  h^{1/4}  m \bar{\phi} \phi.
			\end{eqnarray}
		\end{subequations}
	\end{widetext}

	The reader can check that under the choice of the parameters given in Sec.~\ref{choice1}, the Hamiltonian constraint becomes
	\begin{eqnarray}
		\tilde{\tilde{\mathcal{H}}} & = & - \kappa \epsilon_{ijk} \tilde{\Pi}^{ai} \tilde{\Pi}^{bj} R_{ab}{}^{k} + 2 \kappa \tilde{\Pi}^{a[i} \tilde{\Pi}^{|b|j]} \Psi_{ai} \Psi_{bj} \notag \\
		&& + \dfrac{1}{2}  \tilde{\Pi}^{ai} \left( \bar{\phi} \gamma_{i} \nabla_{a} \phi - \overline{\nabla_{a} \phi} \gamma_{i} \phi \right) \notag \\
		&&+ \dfrac{n_{0}}{2}  \epsilon_{ijk} \Psi_{a}{}^{i}  \tilde{\Pi}^{aj} \tilde{A}^{k} + \dfrac{3}{32 \kappa}\left( \tilde{A}^{0}\right)^{2}    \notag \\
		\label{const_Psi_H_tg_A}
		& &  - 2 h^{1/2}  \kappa \Lambda  -  h^{1/4} m \bar{\phi} \phi.
	\end{eqnarray}
	Similarly, under the choice of the parameters given in Sec.~\ref{choice2}, the Hamiltonian constraint becomes
	\begin{eqnarray}
		\tilde{\tilde{\mathcal{H}}} & = & - \kappa \epsilon_{ijk} \tilde{\Pi}^{ai} \tilde{\Pi}^{bj} R_{ab}{}^{k} + 2 \kappa \tilde{\Pi}^{a[i} \tilde{\Pi}^{|b|j]} \Psi_{ai} \Psi_{bj} \notag \\
		&& + \dfrac{1}{2}  \tilde{\Pi}^{ai} \left( \bar{\phi} \gamma_{i} \nabla_{a} \phi - \overline{\nabla_{a} \phi} \gamma_{i} \phi \right) \notag \\
		&&+ \dfrac{n_{0}}{2}  \epsilon_{ijk} \Psi_{a}{}^{i}  \tilde{\Pi}^{aj} \tilde{A}^{k} + \dfrac{3}{32 \kappa}\Big[ \vartheta^{2} \tilde{V}_{I} \tilde{V}^{I} + \left(\tilde{A}^{0}\right)^{2}   \Big] \notag \\
		\label{const_Psi_H_tg_B}
		& &  - 2 h^{1/2}  \kappa \Lambda  -  h^{1/4} m \bar{\phi} \phi.
	\end{eqnarray}

	\section{Conclusions}\label{Sec_concl}
	Our first conclusion is that the approach of Ref.~\cite{Montesinos2004a} is robust enough because it has 
	allowed us to get the Hamiltonian formulation involving only first-class constraints of minimally and nonminimally coupled fermion fields to the Holst action leaving intact the local Lorentz symmetry, all of this by means of first a suitable parametrization of the frame and the connection and later by integrating out the auxiliary fields. We recall that local Lorentz symmetry is what allows the coupling of fermion fields to first-order general relativity at the Lagrangian level, and so it is remarkable that the Hamiltonian analysis of the theory can be carried out without spoiling this fundamental symmetry of nature. We consider this fact one of the relevant results of this paper. From it we deduce that the Hamiltonian formulation of the coupling of fermion fields to the $n$-dimensional Palatini action can also be carried out by following the same approach reported in Ref.~\cite{Montesinos2001} (work is in progress~\cite{MRC_PalFer}). Furthermore, regarding the manifestly Lorentz-invariant Hamiltonian formulations reported in Secs.~\ref{Sec_Ham_form}--\ref{Sec_hf2}, we consider them relevant enough in their own right because their respective symplectic structures are real and the phase-space variables for the gravitational sector are also real, although the Hamiltonian formulations that employ half-densitized fermion fields reported in Secs.~\ref{Sec_hf} and~\ref{Sec_hf2} involve simpler expressions for the first-class constraints than the corresponding one of Sec.~\ref{Sec_Ham_form}. We consider the three of them at the same level classically. We also consider remarkable the fact that it is possible to perform a symplectomorphism from the Hamiltonian formulation contained in Sec.~\ref{Sec_hf2}, which does not involve the coupling parameters of fermion fields in the symplectic structure, and that the resulting Hamiltonian formulation contained in Sec.~\ref{Sec_CT} involves simpler expressions for the first-class constraints than that of Sec.~\ref{Sec_hf2}. This fact allows us to get immediately the manifestly Lorentz-invariant Hamiltonian formulation of fermion fields minimally coupled to the Palatini action when a particular choice of the parameters is made. Additionally, the Hamiltonian formulation of fermion fields nonminimally coupled to the Palatini action that is invariant under parity transformations is 
	easily obtained when another particular choice of the parameters is made. Hopefully, our findings might motivate the research in quantum gravity. In particular, the Hamiltonian formulations obtained once the time gauge is imposed on might be relevant for researchers working in loop quantum gravity~\cite{RovBook,ThieBook,Ashtekar0407}.
	
	\acknowledgments
	This work was partially supported by Fondo SEP-Cinvestav and by Consejo Nacional de Ciencia y Tecnología (CONACyT), M\'{e}xico, Grant No. A1-S-7701. M.~C. gratefully acknowledges the support of a DGAPA-UNAM postdoctoral fellowship.

	\appendix

	\section*{Appendix: Maps and algebraic relations}
	\label{Appendix}
	We follow the results of Ref.~\cite{Montesinos2004a}. The map $(N, N^a, \tilde{\Pi}^{aI}) \mapsto (e_{\mu}{}^I)$ given by~\eqref{et_rp} and~\eqref{ea_rp} has the inverse map $(e_{\mu}{}^I) \mapsto (N, N^a, \tilde{\Pi}^{aI})$,
	\begin{subequations}
		\begin{eqnarray}
			N &=& - n_{I} e_{t}{}^{I},\\
			N^a &=& q^{ab} e_{t}{}^{I} e_{bI},\\
			\tilde{\Pi}^{aI} &=& \sqrt{q} q^{ab} e_{b}{}^{I},
		\end{eqnarray}
	\end{subequations}
	where $q^{ab}$ is the inverse of the induced metric on $\Sigma$ ($q_{ab}=e_{a}{}^{I} e_{bI}$) and $q=\det (q_{ab})$. On the other hand, $n_{I}$ is given in terms of $e_{a}{}^{I}$ as
	\begin{equation}
		n_{I} = \dfrac{1}{6 \sqrt{q}} \epsilon_{IJKL} \tilde{\eta}^{abc} e_{a}{}^{J} e_{b}{}^{K} e_{c}{}^{L}.
	\end{equation}
	From this expression, it is clear that $n_{I}$ is a normal vector to the hypersurface $\Sigma$.
	
	The map $(C_{aI}, \uac{\lambda}_{ab}, \lambda_{IJ}) \mapsto (\omega_{\mu}{}^I{}_J)$ given by~\eqref{w} and \eqref{l} has the inverse map  $(\omega_{\mu}{}^{I}{}_{J}) \mapsto (C_{aI}, \uac{\lambda}_{ab}, \lambda_{IJ})$ composed by~\eqref{C_def} and
	\begin{eqnarray}
		\uac{\lambda}_{ab} &=& \uac{U}_{ab}{}^{cIJ} \ovg{\omega}_{cIJ}, \\
		\lambda_{IJ} & = & -\omega_{tIJ} + N^{a} \omega_{aIJ} - 2 \tilde{\Pi}^{a}{}_{[I} n_{J]} \nabla_{a} \uac{N} \notag \\
		& &  - \dfrac{1}{4 \kappa} \uac{N} \bigg\lbrace \tilde{\mathcal{G}}_{IJ} - \left( P^{-1} \right)_{IJKL} \tilde{\mathcal{G}}^{KL} \notag \\
		& & - 2 n^{K} n_{[I}  \tilde{\mathcal{G}}_{J]K}   - h^{1/4} \bigg[  \ast \left( P^{-1} \right)_{IJKL} n^{K} A^{L}  \notag \\
		& & + \left( P^{-1} \right)_{IJKL} n^{K}  \big( \theta V^{L} + \xi A^{L} \big)   \notag \\
		& &   - \dfrac{1}{2} \epsilon_{IJKL}  n^{K} A^{L}  \bigg] \bigg\rbrace,
	\end{eqnarray}
	where $\uac{U}_{ab}{}^{cIJ}$ is defined back in Eq.~\eqref{U}. The geometrical objects (projectors) involved in these maps $M_{a}{}^{b}{}_{IJK} = - M_{a}{}^{b}{}_{JIK}$, $\tilde{N}^{a}{}_{IJ} = - \tilde{N}^{a}{}_{JI}$, $W_{a}{}^{b}{}_{IJK} = - W_{a}{}^{b}{}_{IKJ}$, and  $\uac{U}_{ab}{}^{cIJ} = \uac{U}_{ba}{}^{cIJ} = -\uac{U}_{ab}{}^{cJI}$ satisfy the orthogonality relations
	\begin{subequations}
		\begin{eqnarray}
			W_{a}{}^{cIKL} M_{c}{}^{b}{}_{KLJ} & = & \delta_{a}^{b} \delta_{J}^{I},\\
			\uac{U}_{ab}{}^{cIJ} \tilde{N}^{d}{}_{IJ} & = & \delta^{c}_{(a} \delta^{d}_{b)},\\
			W_{a}{}^{(b}{}_{IJK}\tilde{N}^{c)JK} & = &0,\\
			\uac{U}_{ab}{}^{cIJ} M_{c}{}^{d}{}_{IJK} & = &0,
		\end{eqnarray}
	\end{subequations}
	and the completeness relation
	\begin{equation}
		M_{a}{}^{c}{}_{IJM} W_{c}{}^{bMKL} + \tilde{N}^{c}{}_{IJ} \uac{U}_{ac}{}^{bKL} = \delta_{a}^{b} \delta_{[I}^{K} \delta_{J]}^{L}.
	\end{equation}

	\bibliographystyle{apsrev4-1}
	
	\bibliography{References}
	
\end{document}